\documentclass[traditabstract]{aa} 
\usepackage{graphicx}
\usepackage{natbib}
\usepackage{amssymb,amsmath}
\usepackage{txfonts}
\begin{document}
\newcommand{\Zsolar}{\mbox{$\,\rm Z_{\odot}$}}
\newcommand{\Msolar}{\mbox{$\,\rm M_{\odot}$}}
\newcommand{\Lsolar}{\mbox{$\,\rm L_{\odot}$}}
\newcommand{\xs}{$\chi^{2}$}
\newcommand{\dxs}{$\Delta\chi^{2}$}
\newcommand{\xsn}{$\chi^{2}_{\nu}$}
\newcommand{\ls}{{\tiny \( \stackrel{<}{\sim}\)}}
\newcommand{\gs}{{\tiny \( \stackrel{>}{\sim}\)}}
\newcommand{\asec}{$^{\prime\prime}$}
\newcommand{\amin}{$^{\prime}$}
\newcommand{\mstar}{\mbox{$M_{*}$}}
\newcommand{\hi}{H{\sc i}}
\newcommand{\hii}{H{\sc ii}\ }
\newcommand{\kms}{$\rm km~s^{-1}$}

   \title{The dust scaling relations of the {\it Herschel} Reference Survey\thanks{Herschel is an ESA space observatory with science 
   instruments provided by European-led Principal Investigator consortia and with important participation from NASA.}}

\author{
L. Cortese\inst{1} 
\and
L. Ciesla\inst{2} 
\and
A. Boselli\inst{2}
\and
S. Bianchi\inst{3}
\and
H. Gomez\inst{4}
\and
M. W. L. Smith\inst{4}
\and 
G. J. Bendo\inst{5}
\and
S. Eales\inst{4}
\and
M. Pohlen\inst{4}
\and
M. Baes\inst{6} 
\and
E. Corbelli\inst{3}
\and
J. I. Davies\inst{4}
\and
T. M. Hughes\inst{7}
\and
L. K. Hunt\inst{3}
\and
S. C. Madden\inst{8}
\and
D. Pierini\inst{9}
\and
S. di Serego Alighieri\inst{3}
\and
S. Zibetti\inst{10}
\and
M. Boquien\inst{2}
\and
D. L. Clements\inst{11}
\and
A. Cooray\inst{12}
\and
M. Galametz\inst{13}
\and
L. Magrini\inst{3}
\and
C. Pappalardo\inst{3}
\and 
L. Spinoglio\inst{14}
\and
C. Vlahakis\inst{15,16} 
%
}

\institute{
European Southern Observatory, Karl Schwarzschild Str. 2, 85748 Garching bei M\"unchen, Germany
\and
Laboratoire d'Astrophysique de Marseille, UMR 6110 CNRS, 38 rue F. Joliot-Curie, F-13388 Marseille, France 
\and
INAF-Osservatorio Astrofisico di Arcetri, Largo Enrico Fermi 5, 50125 Firenze, Italy 
\and
School of Physics and Astronomy, Cardiff University, The Parade, Cardiff, CF24 3AA, UK
\and
UK ALMA Regional Centre Node, Jodrell Bank Centre for Astrophysics, School of Physics and Astronomy, University of Manchester, Oxford Road,
Manchester M13 9PL, United Kingdom
\and
Sterrenkundig Observatorium, Universiteit Gent, Krijgslaan 281 S9, B-9000 Gent, Belgium 
\and
Kavli Institute for Astronomy \& Astrophysics, Peking University, Beijing 100871, China
\and
Institut d'Astrophysique Spatiale (IAS), Batiment 121, Universite Paris-Sud 11 and CNRS, F-91405 Orsay, France 
\and
Visiting Astronomer, Max-Planck-Institut f\"ur extraterrestrische Physik, Giessenbachstrasse, Postfach 1312, D-85741, Garching bei M\"unchen, Germany
\and
Dark Cosmology Centre, Niels Bohr Institute University of Copenhagen, Juliane Maries Vej 30, DK-2100 Copenhagen, Denmark
\and
Astrophysics Group, Blackett Lab, Imperial College, Prince Consort Road, London SW7 2AZ, UK
\and
University of California, Irvine, Department of Physics \&  Astronomy, 4186 Frederick Reines Hall, Irvine, CA, USA
\and
Institute of Astronomy, University of Cambridge, Madingley Road, Cambridge, CB3 0HA, UK
\and
Istituto di Fisica dello Spazio Interplanetario, INAF, Via Fosso del Cavaliere 100, I-00133 Roma, Italy
\and
Joint ALMA Office, Alonso de Cordova 3107, Vitacura, Santiago, Chile
\and
Departamento de Astronomia, Universidad de Chile, Casilla 36-D, Santiago, Chile
%
}

\date{Received 22 November 2011 - Accepted 10 January 2012}

 
  \abstract{We combine new {\it Herschel}/SPIRE sub-millimeter observations with existing multiwavelength data to investigate the dust 
  scaling relations of the Herschel Reference Survey, a magnitude-, volume-limited sample of $\sim$300 nearby galaxies 
  in different environments. We show that the dust-to-stellar mass ratio anti-correlates with stellar mass, stellar 
  mass surface density and $NUV-r$ colour across the whole range of parameters covered by our sample.
  Moreover, the dust-to-stellar mass ratio decreases significantly when moving from late- to early-type galaxies. 
  These scaling relations are similar to those observed for the \hi\ gas-fraction, supporting the idea that 
  the cold dust is tightly coupled to the cold atomic gas component in the interstellar medium. 
  We also find a weak increase of the dust-to-\hi\ mass ratio with stellar mass and colour but no trend is 
  seen with stellar mass surface density. By comparing galaxies in different environments we show that, 
  although these scaling relations are followed by both cluster and field galaxies, \hi-deficient systems have, 
  at fixed stellar mass, stellar mass surface density and morphological type systematically lower dust-to-stellar mass 
  and higher dust-to-\hi\ mass ratios than \hi-normal/field galaxies. This provides clear evidence that dust is removed from 
  the star-forming disk of cluster galaxies but the effect of the environment is less strong than what is observed in the 
  case of the \hi\ disk. Such effects naturally arise if the dust disk is less extended than the \hi\ and follows more closely 
  the distribution of the molecular gas phase, i.e., if the dust-to-atomic gas ratio monotonically decreases with distance 
  from the galactic center. 
  }


	\authorrunning{Cortese et al.}	
	\titlerunning{}
   \maketitle
%

\section{Introduction}
One of the main challenges for extragalactic astronomy is to understand 
how galaxies evolved from simple clouds of un-enriched gas into the complex eco-systems 
harboring stars, planets, dust, metals and different gas phases we observe today.
Of course, this {\it transformation} must be related to the 
star formation cycle in galaxies. The gas cools, condensing into molecular clouds that 
collapse forming stars. Heavy elements are formed during nuclear reactions in the stellar cores and then, 
when the stars die, they are expelled into the inter-stellar medium (ISM) and either mix 
with the gas phase or condense forming dust grains. 
However, we still know very little about the detailed physical processes regulating 
the star formation cycle in galaxies and how much it is affected by the internal 
properties of galaxies (e.g., mass, dynamics and morphology) and the environment 
they inhabit.

In the last decades, wide-area optical surveys of the local Universe and 
deep investigations of high-redshift systems have allowed an accurate reconstruction 
of the average star formation history (SFH) of galaxies in different environments 
across the Hubble time (e.g., \citealp{sdssDR7,cosmos}). They have not only revealed that the star 
formation rate (SFR) density of the Universe had a peak around $z\sim$1 and is now 
rapidly declining (e.g., \citealp{lilly96,madau98}), but they have also shown that the SFH of galaxies 
is tightly linked to their total mass (i.e., massive systems 
formed the bulk of their stellar populations earlier than dwarf galaxies; \citealp{cowie96,phenomen}), 
and to the environment (i.e., at the current epoch, galaxies in high density regions are less active 
than isolated systems; \citealp{lewis02,gomez03,review}). 
The next step is to investigate why the conversion of gas into stars is mass 
and environment dependent and how this impacts on the metal enrichment of the 
ISM. 


A unique role in the star formation cycle of galaxies is played by dust grains.
Since dust is formed out of the metals produced during stellar nucleosynthesis, its 
properties provide us with important clues about the recent star formation activity of galaxies. 
Moreover, dust is also believed to act as catalyzer for the formation of molecular hydrogen 
(H$_{2}$) and prevents its dissociation by the interstellar radiation field, thus helping 
to regulate the future star formation activity (e.g., \citealp{hollenbach71}). Finally, since dust can also be destroyed, 
in particular if not embedded into the cold ISM (e.g., \citealp{clemens10}), 
its presence and abundance can yield important information about the properties of the other components 
of the ISM in galaxies.

Although dust properties have been studied for several decades \citep{savage79}, it has become gradually clear 
that, in order to obtain a proper census of dust in the Universe, it is necessary to incorporate  
extragalactic investigations of the far-infrared/sub-millimeter (submm) regime. 
It is in this part of the electromagnetic spectrum that the cold dust component 
dominating the dust mass budget in galaxies hides. Pioneers in this area have been 
the IRAS \citep{iras}, ISO \citep{iso} and {\it Spitzer} \citep{spitzer} telescopes and the SCUBA camera \citep{scuba}, 
but now with the launch of {\it Herschel} \citep{pilbratt10} are we entering a new era for interstellar dust studies. 
Thanks to its unprecedented sensitivity in the wavelength range 200$<\lambda<$600 $\mu$m and a much improved 
spatial resolution at 70$<\lambda<$160 $\mu$m (with respect to {\it Spitzer}),
{\it Herschel} is a unique instrument to unveil the role played by dust in the evolutionary history 
of galaxies. 

The first, natural step in this direction is to quantify how the dust content of galaxies varies 
with internal galaxy properties such as, stellar mass, colour, surface density, gas content, etc. 
These {\it scaling relations} will provide initial clues on the role played by dust on the star formation 
cycle and strong constraints for chemical evolution models. For example, it is 
important to quantify the link between the dust and the cold gas component of the ISM and investigate 
whether or not they follow similar scaling relations. 

The main scaling relation investigated during the pre-{\it Herschel} era has been the evolution of the dust-to-gas 
ratio with stellar mass and metallicity (e.g., \citealp{issa90,lisenfeld98,popescu02,draine07,galametz11}). 
Several studies have shown an increase of the dust-to-gas ratio as a function of stellar mass, mimicking what 
has already been observed in the case of the stellar mass vs. gas-phase metallicity relation \citep{tremonti04}.
Although several investigations have quantified the variation 
of the total far-infrared-to-optical/near-infrared luminosity (usually interpreted as a proxy 
for the ratio between obscured star formation rate and stellar mass) as a function of morphological type 
(e.g., \citealp{dejong84,devereux97,bendo02,popescu02}), 
surprisingly very little is known about the relation between the dust-to-stellar mass ratio and galaxy properties. 
Only very recently, \cite{dacunha10} have shown that the dust-to-stellar mass 
strongly correlates with specific star formation rate (i.e., the ratio between the current 
star formation rate and the stellar mass, SSFR), as predicted by chemical evolution models. 
However, it is still unknown whether or not morphology and stellar mass also regulate the amount of 
dust present in nearby galaxies. 

Even more limited is our knowledge of the role (if any) played by the environment 
on the dust content of galaxies. Whereas it is now well established that the atomic 
hydrogen (H{\sc i}) content of galaxies depends on the environment they inhabit, it 
is still debated whether or not the harsh cluster environment can also affect the 
molecular hydrogen component in the ISM \citep{boselligdust,fumagalli09} . 
Thus, since the dust is supposed to be more directly linked to the molecular than to the atomic 
gas phase, it is not clear if dust is removed from the star-forming disks of cluster galaxies. 
Before the launch of {\it Herschel}, no definitive evidence of dust stripping 
in cluster galaxies had been found (e.g., \citealp{tuffs02,popescu02,contursi01,review}). 

The situation has considerably improved in the last year, thanks to the {\it Herschel} 
 Science Demonstration Phase. {\it Herschel} observations  
have not only allowed us to directly witness dust stripping in clusters of galaxies 
\citep{cortese10b,gomez10}, but they have also clearly demonstrated that the dust disk is 
significantly reduced in very H{\sc i}-deficient cluster galaxies, 
following remarkably well the observed `truncation' of the H{\sc i} disk \citep{cortese10c}.
Moreover, \cite{walter11} have found evidence of dust stripping also in the M81 triplet. 
Thus, while it is becoming clear that dust can really be perturbed by the environment, 
it is still uncertain whether these are just extreme cases or dust stripping 
is a common evolutionary phase for cluster galaxies.

Dust scaling relations are the necessary step to solve this issue: 
only after the relation between internal galaxy properties and dust content has 
been properly quantified, will it be possible to determine the role played by the environment. 
It is in fact mandatory to compare the properties of galaxies in different environments 
at fixed stellar mass, morphological type and colour to exclude that our findings 
are just a result of secondary trends between galaxy structure, star formation 
activity and environment. 

For all these reasons, in this paper we take advantage of {\it Herschel} observations 
of the Herschel Reference Survey (HRS, \citealp{HRS}) to quantify how the dust-to-stellar mass 
and dust-to-\hi\ mass vary with internal galaxy properties and environment. 
The HRS is the ideal sample to carry out this analysis in the local Universe. 
Its selection criteria (magnitude- and volume-limited), size ($\sim$300 galaxies), 
sensitivity to the cold dust component (down to $\sim$10$^{4}$ M$_{\odot}$ at the distance 
of the Virgo cluster) and multiwavelength coverage, make this sample ideal to investigate 
dust properties in the local Universe, thus providing strong constraints for theoretical models 
and a reference for high-redshift studies.

This paper is organized as follows. In \S~2 we describe the {\it Herschel} observations and 
data reduction, the technique used to estimate the dust mass and the ancillary data used 
to trace the other components of the ISM. In \S~3 and \S~4 we discuss the relation 
of the dust-to-stellar mass and dust-to-gas mass with galaxy properties and environment while 
in \S~5 we discuss the implications of our results for chemical evolution models and 
for environmental studies. In Appendix we investigate possible biases affecting 
our dust mass estimates.


\section{The data}
The HRS consists of a volume-limited sample 
(i.e., 15$\leq D \leq$25 Mpc) including late-type galaxies (Sa and later) with 
2MASS \citep{2massall} K-band magnitude K$_{Stot} \le$ 12 mag and early-type galaxies (S0a and earlier) with 
K$_{Stot} \le$ 8.7 mag. Additional selection criteria are high galactic latitude 
(b $>$ +55$^{\circ}$) and low Galactic extinction (A$_{B}$ $<$ 0.2 mag, \citealp{schlegel98}), 
to minimize Galactic cirrus contamination.
The total sample consists of 322 galaxies (260 late- and 62 early-type galaxies\footnote{With respect to the original 
selection described in \cite{HRS}, we removed HRS228 because the redshift reported in NED turned out 
to be incorrect. In addition, after visual inspection and comparison with the literature, we revised the NED-based 
morphological type classification given in \cite{HRS} for six galaxies in our sample. 
In detail, NGC4179 moved from Sb to S0, NGC4438 from S0/a to Sb, NGC4457 from S0/a to Sb, VCC1549 from Sb to dE, NGC4691 
from S0/a to Sa and NGC5701 from S0/a to Sa \citep{shapley,vcc,goldmine,cortese09}.}).
As extensively discussed in \cite{HRS}, this sample is not only representative of the local Universe but also spans 
different density regimes (i.e., from isolated systems to the center of the Virgo cluster) and so it is ideal for environmental 
studies (see also \citealp{hughes09,cortese09,boselliimf,cortese11}).
As discussed in \cite{HRS}, we fixed the distances for galaxies belonging to the Virgo cluster (i.e., 23 Mpc for the 
Virgo B cloud and 17 Mpc for all the other clouds; \citealp{gav99}), while for the rest of the sample 
distances have been estimated from their recessional velocities assuming a Hubble constant $H_{0}=$70 km s$^{-1}$ Mpc$^{-1}$.

\subsection{Herschel/SPIRE observations, data reduction and flux extraction}

{\it Herschel}/SPIRE \citep{spire} observations of galaxies in the HRS 
were originally planned as part of the Guaranteed Time of the SPIRE and PACS consortia.
However, after the selection of the {\it Herschel} Open Time Key Projects, it 
emerged that 83 HRS galaxies were also targeted, to a similar depth, by the Herschel Virgo 
Cluster Survey (HeViCS, \citealp{davies10}), a key program focused on the study of  
$\sim$60 deg$^{2}$ in the Virgo cluster using PACS/SPIRE \citep{spire,pacs} parallel mode. 
In order to reduce as much as possible the duplication of {\it Herschel} observations, 
the local galaxies group of the SPIRE science team (SAG2) and the HeViCS team have agreed to share the 
data obtained for the galaxies in common between the two programs. 
Thus, the {\it Herschel}/SPIRE data presented in this work have been obtained as part of these two {\it Herschel} Key Projects. 

The 239 galaxies outside the HeViCS footprint, plus 4 galaxies in HeViCS targeted by the HRS 
during the {\it Herschel} Science Demonstration Phase, have been observed using the SPIRE scan-map 
mode with a nominal scan speed of 30 $^{\prime\prime}$/sec. The size of each map was chosen 
on the basis of the optical extent of the target. Galaxies with optical diameters smaller than 
$\sim$3\arcmin\ were observed using the small scan-map mode providing homogeneous 
coverage over a circular area of $\sim$5\arcmin\ diameter. For larger galaxies, the large scan-map mode 
was used, with each map covering at least an area with diameter 1.5 times the optical diameter 
of the target. The typical sizes of these maps are $8\arcmin\times8\arcmin$, $12\arcmin\times12\arcmin$ and 
$16\arcmin\times16\arcmin$. 
The depth of each map was decided according to the morphological type 
of the target: 3 and 8 pairs of cross-linked scan-map observations for late- and early-type 
galaxies, respectively. This resulted into a pixel-by-pixel rms at 
250, 350 and 500 $\mu$m of $\sim$7, 8, 9 mJy/beam for late-type galaxies and 
$\sim$6, 6 and 7 mJy/beam for early-type systems. For additional details see Ciesla et al. (in prep.).

The remaining 79 HRS galaxies targeted by the HeViCS survey have been observed by {\it Herschel} using the 
SPIRE/PACS parallel scan-map mode with a fast scan speed of 60 $^{\prime\prime}$/sec.
Eight cross-scans fields of $\sim 4\times 4$ deg$^{2}$ each across the Virgo cluster were observed 
covering a total area of $\sim$60 deg$^{2}$ (see \citealp{davies10,davies11} for additional details).
The typical pixel-by-pixel rms measured around the HRS galaxies are $\sim$7, 7 and 8 mJy/beam at 
250, 350 and 500 $\mu$m. 

Both datasets were reduced using the same data reduction technique.
Extensive details of our data reduction procedure can be found in 
Smith et al. (in prep.). 
In the following we briefly summarize the main 
steps. The SPIRE photometer data were processed up to Level-1 
with a script adapted from the official SPIRE pipeline.
The main differences with respect to the standard pipeline are 
that we use a different deglitching method and did not run 
the default temperature drift correction and the residual baseline 
subtraction. Instead, we used the BriGAdE method (Smith et al. in
prep.) to remove the temperature drift and bring all bolometers
to the same level (equivalent to baseline removal). 
This method provides a better baseline subtraction, in particular 
when strong temperature variations are present. 
All scans were then combined into a map using the naive map maker 
included in the standard pipeline. 
The pixel sizes of each map are 6, 8 and 12 arcsec and the 
corresponding full widths half maximum (FWHM) 18.2\arcsec, 
24.5\arcsec, 36.0\arcsec\ at 250, 350, 500 $\mu$m, respectively. 

Total integrated flux densities were obtained as extensively described in 
Ciesla et al. (in prep.). Briefly, for extended sources\footnote{We define 
a source as `extended' if a Gaussian fit to the time-line data provides a FWHM larger than 
20\arcsec, 29\arcsec, 37\arcsec\ at 250, 350, 500 $\mu$m, respectively.} elliptical apertures with a typical major 
axis 1.4 times the optical diameter of the galaxies were used.
In some cases (e.g., early-type galaxies) the size of the aperture was significantly reduced in order to 
include all the emission from the galaxy and/or avoid as much as possible contamination 
from other sources or foreground cirrus emission.  
Background was estimated using circular annuli around the target (Ciesla et al., in prep.). Errors were estimated following \cite{bosiso03}. 
For point-like sources (8 out of 322 galaxies at 500$\mu$m), flux densities were extracted directly 
from the time-line data using the method developed by Bendo et al. (in prep.). This turns 
out to provide a more reliable flux density estimate than the aperture photometry 
technique. Errors for point sources come directly from the time-line fitting technique. 
The typical errors on the flux estimates for all HRS galaxies are $\sim$6\%, 8\% and 11\% at 
250, 350, 500 $\mu$m, respectively.
These do not include the SPIRE calibration error of $\sim$7\%.
In case of non-detections, upper-limits have been determined assuming a 3$\sigma$ signal over 
a circular aperture of radius 0.3, 0.8 and 1.4 the optical radius for E, S0 and spirals, respectively (Ciesla et al., in prep.). 
In the following, we will work with Relative Spectral Responsivity Function (RSRF)-weighted 
flux density measurements. Since the SPIRE pipeline produces monochromatic flux densities 
assuming that the source is point-like, we had to apply correction factors to convert 
monochromatic flux densities into RSRF-weighted flux densities. Thus, 
we divided monochromatic flux densities obtained from the maps by 1.0113, 1.0087 and 1.0065, 
at 250, 350 and 500$\mu$m, respectively (see \citealp{spireman}\footnote{http://herschel.esac.esa.int/Docs/SPIRE/html/spire\_om.html}, 
and the SPIRE Photometer Cookbook\footnote{http://herschel.esac.esa.int/twiki/pub/Public/SpireCalibrationWeb/\\SPIREPhotometryCookbook\_may2011\_1.pdf}).   

In total, 276 out of the 322 HRS galaxies ($\sim$86\%) have been detected in all the three 
SPIRE bands, 38 galaxies are non-detections in all bands, 3 have been detected at 250$\mu$m only 
and 5 galaxies at 250 and 350$\mu$m only.

\subsection{Dust mass estimate}
\label{dustsec}
The main aim of this paper is to investigate the relation between dust content and internal and 
environmental properties of galaxies. 
Thus, an accurate estimate of dust masses is mandatory. 
Ideally, the best method to quantify the amount of dust in galaxies would be 
via spectral energy distribution (SED) fitting of data across the whole 
ultraviolet to far-infrared/submm regime using up-to-date dust models 
and energy balance/radiative transfer techniques (e.g., \citealp{zubko04,draine07b,bianchi08,galliano08,dacunha08,baes10,compiegne11}) or just 
assuming simple single/double modified black-body emission (e.g., \citealp{dunne2000,Vlahakis05}). 
Unfortunately, no homogeneous dataset for all the HRS galaxies is currently available for $\lambda<$200 $\mu$m (see \citealp{HRS}), making 
an accurate SED fitting of only the three SPIRE flux densities unfeasible in the case of detailed dust models 
and perhaps not ideal even in the case of a single modified black-body emission \citep{shetty09a,shetty09b}. 
This is mainly because, with only the SPIRE points, no constraints on the shape of the SED below 250 $\mu$m are available 
and no detailed error analysis can be performed. 
Thus, in order to avoid introducing systematic biases due to the incomplete coverage of the HRS 
in the mid- and far-infrared wavelength regime, we derive empirical recipes to estimate dust masses from the three 
SPIRE flux densities only. This approach is similar to what has been done in the past to determine stellar masses from 
optical and near-infrared luminosities and colours (e.g., \citealp{bell03,zibetti09}). 
In recent years, these empirical calibrations have become the ideal tool to roughly quantify 
the stellar content of galaxies when accurate SED fitting is not possible. 
Thus, the technique here adopted must not be considered equivalent to a proper SED fitting, 
but it just provides us with a way to have a first view on the dust properties of the HRS sample.
 
As a first step, we test the feasibility of this technique by using two different approaches. 
Firstly, we assume that the dust SED is well approximated by a simple single modified black-body radiation, 
then we will compare our results with the predictions of the models developed by \cite{draine07b}.
A discussion on the possible biases affecting our dust mass estimates is also presented in Appendix A.

\subsubsection{Modified black-body}
\begin{figure}
\centering
\includegraphics[width=8cm]{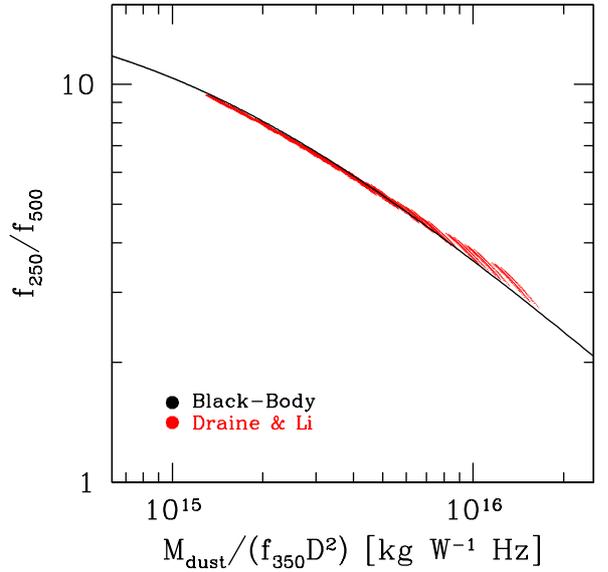}
\caption{The relation between the ratio $M_{dust}/(F_{350} D^{2})$ and the 
250$\mu$m-to-500$\mu$m flux density ratio for a modified black-body SED with $\beta$=2 (black, see Table B1) and for the Draine \& Li models (red).}
\label{dustmod}
\end{figure}

In this case, the dust mass is given by \citep{hildebrand83}
\begin{equation}
M_{dust}= \frac{f_{\nu} D^{2}}{\kappa_{\nu} B_{\nu}(T)}
\label{dustmass}
\end{equation}
where $f_{\nu}$ is the flux density emitted at the frequency $\nu$, $\kappa_{\nu}$ is the dust mass absorption coefficient 
[in the following we will assume  $\kappa_{\nu}=\kappa_{350}(\nu/\nu_{350})^{\beta}$], 
$D$ is the distance and $B_{\nu}(T)$ is the Planck function
\begin{equation}
B_{\nu}(T) = \frac{2 h\nu^{3}}{c^2} \frac{1}{e^{h\nu / KT} - 1}
\end{equation}
where $h$ is the Planck constant, $c$ the speed of light and $T$ is the dust temperature.
We can use Eq.\ref{dustmass} to determine how the 250-to-500$\mu$m flux density ratio varies 
with temperature:
\begin{equation}
\frac{f_{250}}{f_{500}} = \Big(\frac{\nu_{250}}{\nu_{500}}\Big)^{3+\beta} \frac{e^{h\nu_{500} / KT} - 1}{e^{h\nu_{250} / KT} - 1} 
\label{dusttemp}
\end{equation}
and thus, by combining Eq.~\ref{dustmass} and ~\ref{dusttemp}, we can find 
a functional form that relates, for example, the ratio $M_{dust}/(f_{350} D^{2})$ to 
the 250-to-500 $\mu$m ratio. 
We tested this by varying the dust temperature in the range 5-55 K, 
assuming a dust mass absorption coefficient of 0.192 at 350$\mu$m \citep{draine03} and $\beta$=2. 
We choose $\beta$=2 for two reasons. Firstly, a recent analysis of Virgo galaxies (some 
of which in our sample; \citealp{davies11}) has shown that their dust SEDs are well fitted by a single 
modified black-body with $\beta$=2. Secondly, this value is very close to the one assumed in the models by 
\cite{draine07b}, thus simplifying the comparison of our results with more refined dust models (see also \citealp{magrini11}).  
The result of this test is shown in Fig.~\ref{dustmod}, where the black line has been obtained by convolving the black-body SED 
with the SPIRE Relative Spectral Responsivity Function (RSRF) for extended sources.
Unsurprisingly, the two quantities are strongly correlated and it is easy to fit a polynomial function to the data,  
suggesting that this method could be easily used to estimate dust masses when SED fitting is not possible.

\subsubsection{The Draine \& Li (2007) model}
In order to check the validity of our method, we tested it by using a more detailed description 
for the dust SED as given by the dust models developed by \cite{draine07b}\footnote{http://www.astro.princeton.edu/$\sim$draine/dust/irem.html}.
According to these models the dust mass in a galaxy can be related to the flux density 
emitted in the far-infrared/submm by the following equation
\begin{equation}
M_{dust} = \Big(\frac{M_{dust}}{M_H}\Big) m_{H} \frac{F_{\nu} D^{2}}{j_{\nu}}
\label{dustdraine}
\end{equation}
where $M_{dust}/M_H$ is the mass of dust per nucleon (given by the dust model assumed), $m_{H}$ is 
the mass of the hydrogen nucleon and $j_{\nu}$ is the dust emissivity per H nucleon.
As discussed by \cite{draine07b}, the dust emissivity strongly depends on the properties 
of the starlight intensities responsible for the dust heating. In particular, they 
show that the total dust emissivity can be expressed as
\begin{equation}
j_{\nu} = (1-\gamma) j_{\nu}(U_{min}) + \gamma j_{\nu}(U_{min},U_{max}, \alpha)
\label{emdraine}
\end{equation}
where $U_{min}$ and $U_{max}$ are the lowest and highest possible intensities for the 
interstellar radiation field, (1-$\gamma$) is the fraction of dust mass exposed to starlight intensity $U_{min}$ and 
$\alpha$ is the index characterizing the distribution of starlight intensities. 
They find that the SEDs of galaxies in the SINGS sample 
appear to be satisfactorily reproduced with a fixed $\alpha=$2 and $U_{max}=$10$^{6}$, 
so we use these parameters in the rest of our analysis\footnote{We also tested the effects of varying $U_{max}$ and we do not find 
significant differences.}. 
By using the models for a Milky-Way dust (i.e., from MW3.1\_00 to  MW3.1\_60, seven models in total), we 
computed $j_{\nu}$ from Eq.\ref{emdraine} by varying $\gamma$ between 0 and 0.3 (step 0.01) 
and assuming all the possible values for $U_{min}$ (from 0.1 to 25, twenty-three  in total), 
see also \cite{mateos09,wiebe09,willmer09}. Then, we estimated the 250-to-500 $\mu$m flux density ratio and, using Eq.~\ref{dustdraine}, 
the ratio $M_{dust}/(f_{350} D^{2})$ assuming $M_{dust}/M_H=$0.01 \citep{draine07b} and $m_{H}=$1.67$\times$10$^{-27}$ kg.
The red points in Fig.~\ref{dustmod} show the relation between $M_{dust}/(f_{350} D^{2})$ and the 
250$\mu$m-to-500$\mu$m flux density ratio so obtained. 
We recover a relation extremely similar to the one found for the modified black-body case (see also \citealp{magrini11}).
There is almost no systematic offset ($\sim$0.01 dex) between the two methods and the typical scatter between dust mass 
estimates using the two approaches is  $\sim$0.02 dex (see also Appendix A).
We have also compared the results obtained from this method with the dust masses estimated via SED fitting for 
those galaxies in the HeViCS footprint for which PACS 100 and 160$\mu$m are available. We find that the two techniques 
provide consistent results with no systematic effects and a typical scatter of $\sim$0.10 dex (see also Appendix A).

In summary, we have developed a simple method that allows us to estimate dust masses from just the three SPIRE flux densities.
In the rest of the paper we will use the dust masses obtained from the 350$\mu$m flux density and the 250-to-500 $\mu$m ratio 
assuming a modified 
black-body with $\beta$=2. For the non-detections, upper limits to the dust masses are computed 
by using the average $log(f_{250}/f_{500})$ colour observed for the detections (0.75).
Given the uncertainty in each flux density measurement, the typical statistical error in the dust mass is $\sim$0.2 dex 
(not including systematics due to assumptions on the dust properties).
For the estimate of the dust mass, we also treated as non-detections those objects for which the submm emission is 
clearly affected by synchrotron emission: i.e., M84 and M87 (see also \citealp{boselli10,baes10}).
The exact functional forms used for extended and point sources, as well as the one obtained for 
different values of $\beta$ are presented in Appendix B.

\subsubsection{The effect of the dust emissivity on the dust mass estimate}
We have shown that our empirical method is consistent with a typical SED fitting technique, once the properties 
of the underlying dust population are fixed. The \cite{draine07} model is consistent with $\beta$=2 at these wavelengths, so it is not 
completely surprising that we obtain similar results when compared with a modified black-body with $\beta$=2.
Unfortunately, the absolute value of the dust opacity and its dependence on frequency are still quite 
uncertain (e.g., \citealp{fink99,dupac03,gordon10,paradis10,plankbeta}) and it is therefore crucial to investigate how this 
assumption can affect our analysis.
It is easier to address this issue by discussing separately a change in the absolute value of dust opacity 
and in the $\beta$ parameter. A variation in the absolute value of $\kappa_{350}$ would just shift all our dust 
estimates up and down without affecting any correlation (and its significance) that we may find. In other words, this 
will just move systematically the y-axis in all our plots. Much more critical would be a change in the value 
of $\beta$, since this could not only create systematic offsets in the dust mass estimates but also alter 
the shape of the scaling relations by introducing secondary effects with the quantities we are interested in 
(e.g., stellar mass, colour and stellar mass surface density), thus undermining the reliability of our analysis. 
Therefore, in Appendix A, we have investigated how our analysis would be affected by using other  
values of $\beta$: namely, 1, 1.5 and 2.5. 
We find that, in addition to a systematic shift in the dust mass values (up to $\sim$1 dex when moving 
from $\beta$=2.5 to 1), a change of $\beta$ would also introduce some artificial trends, in particular with 
stellar mass and stellar mass surface density. In particular, moving from $\beta$=2.5 to $\beta$=1 dust masses  
would gradually decrease, but galaxies with higher stellar masses and stellar mass surface densities would 
be more ($\sim$0.2 dex) affected than lower mass systems. Luckily, such differences between high and low 
stellar mass systems starts to be important only for $\beta<$1.5. For $\beta\geq$1.5 the scatter in the relations 
is $\leq$0.1 dex, i.e. comparable to, or lower than, the scatter in our method and, as we will see, 
significantly lower than the typical dynamical range covered by the scaling relations we investigate here.
Moreover, it is important to note that the bluest submm colours in our sample (i.e., $f_{250}/f_{500}>$7.5) 
cannot be reproduced by a single modified black-body with $\beta$=1 (see also \citealp{boselli12}).

Thus, we can confidently proceed by estimating the dust masses using the technique developed above 
for a modified black-body with $\beta$=2, and we will discuss throughout the text if and how much different 
values of $\beta$ could influence our conclusions.
Given all the caveats discussed above, we remind the reader to be cautious of the 
absolute values of dust masses presented here, and to mainly focus on the trends and differences between 
the various samples discussed in the rest of the paper.

\begin{figure}
  \centering
  \includegraphics[width=8cm]{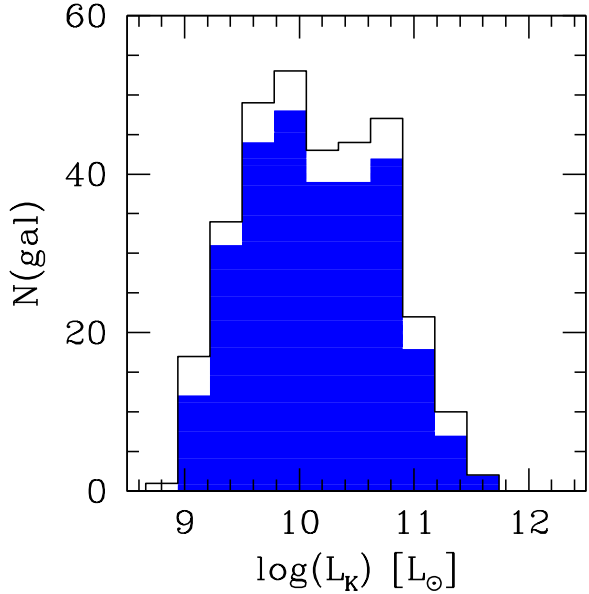}
     \caption{The K-band luminosity distribution for the whole HRS (black) and for the sub-sample analyzed in this paper (blue).}
	 \label{sample}
  \end{figure}

\subsection{Ultraviolet, optical and HI data}
The {\it Herschel}/SPIRE data have been combined with multiwavelength observations covering 
the ultraviolet, optical and \hi\ regime. A detailed description of these datasets and of the 
derived quantities is given in \cite{cortese11}. 
Briefly, optical broad-band photometry has been obtained from the {\it Sloan Digital Sky Survey} 
DR7 (SDSS-DR7, \citealp{sdssDR7}) database. 
{\it Galaxy Evolution Explorer} (GALEX, \citealp{martin05}) near-ultraviolet 
(NUV; $\lambda$=2316 \AA: $\Delta \lambda$=1069 \AA) images have been mainly 
obtained as part of two on-going GALEX Guest Investigator programs 
(GI06-12, P.I. L. Cortese and GI06-01, GALEX Ultraviolet Virgo Cluster Survey, \citealp{guvics}). 
Additional frames have been obtained from the GALEX GR6 public release. 
The SDSS images were registered to the GALEX frames and convolved to the NUV 
resolution. Isophotal ellipses were then fit to each 
image, keeping fixed the center, ellipticity and position angle (generally determined in the $i$-band).
The sky background was determined within rectangular regions around the galaxy and subtracted from the images 
before performing the ellipse fitting.   
Asymptotic magnitudes have been determined from the growth curves obtained following 
the technique described by \cite{atlas2006} and corrected for Galactic extinction 
assuming a \cite{cardelli89} extinction law.
An extensive description of GALEX and SDSS measurements as well as all the fluxes will 
be presented in a future work (Cortese et al. in prep.).  
Stellar masses $M_{*}$ are determined from $i$-band luminosities $L_{i}$ using the $g-i$ colour-dependent 
stellar mass-to-light ratio relation from \cite{zibetti09}, assuming a \cite{chabrier} initial mass function (IMF).
We assume a typical uncertainty of 0.15 dex in the stellar mass estimate. 

Atomic hydrogen masses have been estimated from \hi~21 cm line emission data (mainly single-dish), available from the 
literature (e.g., \citealp{spring05hi,giovanelli07,kent08,goldmine} and the {\it NASA/IPAC Extragalactic Database}, NED).
As described in \cite{cortese11}, we estimate the \hi~deficiency parameter ($Def_{HI}$, defined as the difference, in 
logarithmic units, between the expected H{\sc i} mass for an isolated galaxy with the same morphological type and optical 
diameter of the target and the observed value, following \citealp{haynes}) using coefficients that vary with morphological type. 
These coefficients have been calculated for the following types (see also Table 3 in \citealp{bosellicomaI}): S0a and earlier 
\citep{haynes}, Sa-Sab, Sb, Sbc, Sc \citep{solanes96} and Scd to later types \citep{bosellicomaI}.
It is important to remember that, for morphological types earlier than Sa, the estimate of the expected \hi\ mass 
is highly uncertain (see also \citealp{cortese11}). 
In the following, we will consider as `\hi-deficient' galaxies those objects with $Def_{HI}\geq0.5$ (i.e., 
galaxies with 70\% less hydrogen than isolated systems with the same diameter and morphological type). 
\begin{figure*}
  \centering
  \includegraphics[width=14.cm]{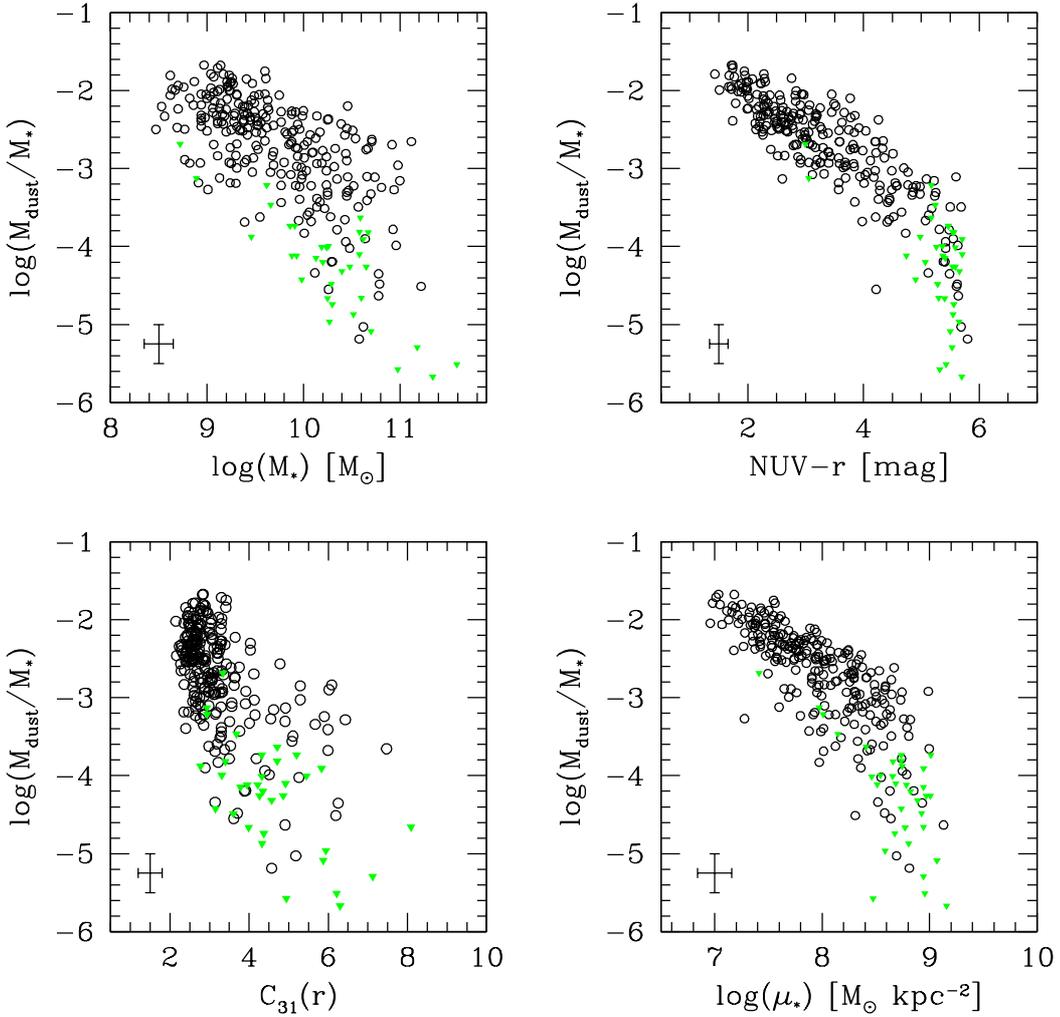}
     \caption{The dust-to-stellar mass ratio as function of stellar mass (upper left), $NUV-r$ colour (upper right), 
     concentration index (lower left) and stellar mass surface density (lower right). Circles and triangles show {\it Herschel} detections 
     and non-detections, respectively. The typical errors associated with our galaxies are indicated on the bottom-left corner 
     of each panel.}
	 \label{MDSall}
  \end{figure*}

In order to avoid any possible bias in the comparison among the various scaling relation, in this paper we focus 
our attention only on those galaxies for which {\it Herschel} as well as \hi, NUV and SDSS observations are currently available: 
282 galaxies ($\sim$87\% of the whole HRS, namely 234 late- and 48 early-type galaxies).
In Fig.~\ref{sample}, we compare the K-band luminosity distribution of the whole HRS (black) and the one of the sub-sample 
here investigated (blue). It is clear that our sub-sample provides a fair representation of the HRS and therefore 
the scaling relations investigated in the rest of the paper should be representative of the local galaxy population as a whole.

\begin{figure*}
  \centering
  \includegraphics[width=18.1cm]{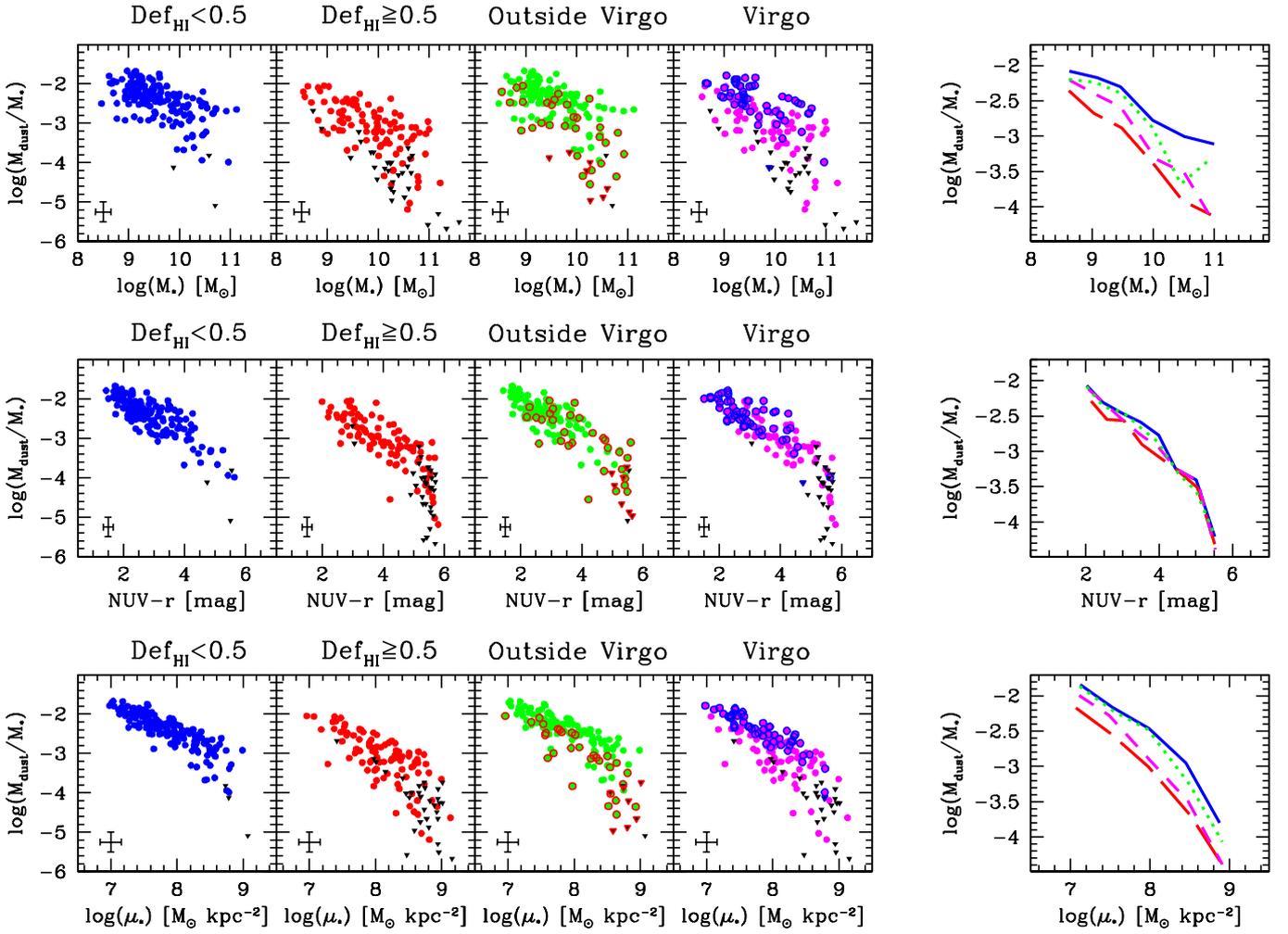}
     \caption{The  dust-to-stellar mass ratio as function of stellar mass (top row), $NUV-r$ colour (middle row), 
    and stellar mass surface density (bottom row) for HI-normal (first column), HI-deficient (second column) galaxies, 
    systems outside (third column) and inside (fourth column) the Virgo cluster. Circles and black triangles indicate detections and non-detections 
    respectively. The average relations for HI-normal (solid), HI-deficient (long-dashed) galaxies, systems outside (dotted) and inside 
    (short dashed) the Virgo cluster are presented in the fifth column. In the third and fourth column \hi-deficient galaxies outside the 
    Virgo cluster and \hi-normal Virgo galaxies are highlighted in red and blue, respectively.\newline}
	 \label{MDSenv}
  \end{figure*}

\section{The dust-to-stellar mass ratio}    
As a first step in order to quantify how the dust content varies as a function of integrated galaxy properties, 
we plot in Fig.\ref{MDSall} the dust-to-stellar mass ratio as a function of stellar mass $M_{*}$ (upper-left panel), observed 
$NUV-r$ colour (upper-right), concentration index in $r$-band [$C_{31}(r)$, defined as the ratio between the radii 
containing 75\% and 25\% of the total $r$-band light\footnote{Concentration indexes are not corrected for inclination but 
this does not significantly affect the results presented here.}] and stellar mass surface density $\mu_{*}$ [i.e., $M_{*}/(2\pi R_{50,i}^{2})$, 
where $R_{50,i}$ is the radius containing 50\% of the total $i$-band light].
Circles and triangles indicate detections and upper-limits, respectively. 
The dust-to-stellar mass ratio strongly anti-correlates with $M_{*}$, $NUV-r$ 
(a proxy for specific star formation rate, SSFR; e.g., \citealp{schiminovich07}) and $\mu_{*}$, while a very weak non-linear trend is observed 
with the concentration index. The tightest correlation is with the observed $NUV-r$ colour 
(Pearson correlation coefficient $r=-$0.85, dispersion along the y-axis 
$\sigma=$0.38 dex)\footnote{These parameters are obtained using upper limits for the non-detections.}, 
while the scatter gradually increases for the stellar surface density ($r=-$0.79, $\sigma=$0.45 dex) and stellar mass 
($r=-$0.68, $\sigma=$0.53 dex).
These results are remarkably similar to the scaling relation involving the HI-to-stellar mass ratio
\citep{cortese11,catinella10,fabello10}, suggesting that the dust and atomic hydrogen content of galaxies might be 
directly linked. 

As discussed in the previous section, it is important to investigate if and how the assumptions made  
on the dust properties of our sample can affect the shape and significance of the scaling relations. 
In Appendix A we show that only for $\beta<$1.5 systematic effects start to be significant (i.e., $\sim$0.12 dex). 
However, when moving from $\beta$=2 to $\beta<$1.5, the decrease in the value of the dust mass would be larger 
for massive, high stellar mass surface density galaxies, effectively reinforcing the strong 
correlations between the dust-to-stellar mass ratio and stellar mass, stellar mass surface density and 
$NUV-r$ colours. Thus we conclude that the main trends shown in Fig.~\ref{MDSall} are independent 
of the assumptions on the dust properties of the HRS.

Recently, \cite{cortese11} have shown that the HI scaling relations for the whole HRS are slightly biased 
towards lower gas content, with respect to the average scaling relation of local galaxies (e.g., \citealp{catinella10,fabello10}). 
This is because, by construction, nearly half of the galaxies 
in the HRS belong to the Virgo cluster. In order to test whether this is also the case for the dust 
scaling relations, and to characterize the relation between dust-to-stellar mass ratio in different environments, 
we have divided our sample into four different subsets:  
(a) \hi-normal (i.e., $Def_{HI}<0.5$, 158 galaxies) and (b) \hi-deficient (i.e., $Def_{HI}\geq 0.5$, 124 galaxies) systems, 
(c) galaxies outside the Virgo cluster (144 galaxies)  and (d) galaxies belonging to one of the Virgo cluster clouds 
(Virgo A, B, N, E and S as defined by \citealp{gav99}, 138 galaxies).
The two pairs (a)-(b) and (c)-(d) are mutually exclusive, while the other combinations 
are not; thus these are complementary sub-samples. 
The scaling relations for the four samples are plotted in Fig.~\ref{MDSenv}.
In order to properly quantify the difference between different environments, in the right most panel of Fig.~\ref{MDSenv} 
and in Table~\ref{scaletab} we present the average trends (i.e., $<log(M_{dust}/M_{*})>$) for each subsample, determined 
by placing the non-detections to their upper-limit.  
Although for all four samples the dust-to-stellar mass ratio decreases with stellar mass, colour and stellar mass 
surface density, galaxies in different environments have different dust contents.
In particular, for fixed stellar mass and stellar mass surface density, 
Virgo and \hi-deficient galaxies have, on average, a lower dust-to-stellar mass ratio than \hi-normal/field galaxies. 
This difference is particularly strong between \hi-normal and \hi-deficient galaxies (i.e., $\sim$0.5-0.7 dex), while 
it is less remarkable (i.e., $\sim$0.2-0.4 dex) when galaxies are separated accordingly to the environment they inhabit, 
suggesting that the atomic hydrogen content is more important than the local environment in regulating the positions 
of galaxies in the scaling relations. 

In order to better quantify the difference between the various subsamples, we estimated the residuals of the \hi-deficient 
and Virgo galaxies from the mean trends observed for \hi-normal and galaxies outside Virgo, respectively. A simple $\chi^{2}$ test 
indicates that \hi-deficient and \hi-normal galaxies do not follow the same $M_{dust}/M_{*}$ vs. $M_{*}$ and $M_{dust}/M_{*}$ vs. $\mu_{*}$ 
relations at $>$99.99\% significance level. Galaxies in and outside Virgo do not follow the same 
$M_{dust}/M_{*}$ vs. $M_{*}$ and $M_{dust}/M_{*}$ vs. $\mu_{*}$ relations at a $\sim$99.7\% and $\sim$98\% level, respectively. 
A similar conclusion is reached if, instead of performing a $\chi^{2}$, we just compare 
the median residuals of the two populations: while \hi-deficient and \hi-normal galaxies differ by $\sim$7-8 $\sigma$, 
galaxies inside and outside the Virgo cluster show just a $\sim$2-3 $\sigma$ difference.      
The shift towards lower dust content for \hi-deficient/Virgo galaxies is also confirmed 
by the fact that the vast majority of non-detections are found among gas-poor cluster galaxies.   
Finally, also the scatter of the $M_{dust}/M_{*}$ vs. $M_{*}$ and $M_{dust}/M_{*}$ vs. $\mu_{*}$ relations varies among the four sub-samples 
here considered: from $\sim$0.4-0.3 dex to $\sim$0.6-0.5 dex when moving from \hi-normal to \hi-deficient systems and 
from $\sim$0.57-0.44 dex to $\sim$0.63-0.5 dex when moving from objects outside and inside Virgo, respectively. 

\begin{figure}
  \centering
  \includegraphics[width=9.cm]{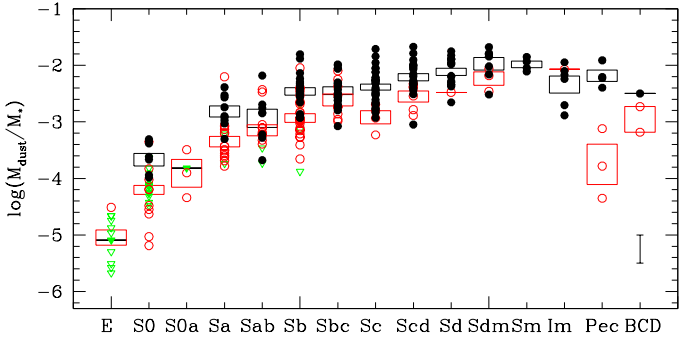}
     \caption{The $M_{dust}/M_{*}$ as a function of the morphological type. Black filled and red empty circles are 
     HI-normal and HI-deficient galaxies, respectively. Upper limits are indicated by green triangles. The large empty rectangles  
     indicate the average value and its error for HI-normal (black) and HI-deficient (red) galaxies in each morphological type. 
     In case only one galaxy is available in a bin, only a line is shown.}
	 \label{MDStypes}
  \end{figure}

It is quite easy to understand why a rough cut in environment is less powerful in isolating 
dust poor systems than a cut in gas content. On one side, our Virgo sample still includes \hi-normal star-forming galaxies 
not yet affected by the cluster environment \citep{review,cortese09}. On the other side, we find \hi-deficient galaxies 
also outside Virgo \citep{cortese11}. These two factors together reduce the difference between the two samples. 
This is clearly shown in Fig.~\ref{MDSenv} where \hi-deficient field galaxies and \hi-normal Virgo galaxies are highlighted. 
It is also important to remember that, due to the lack of statistics available outside the 
Virgo cluster (in particular for early-type galaxies), it is impossible to perform a more detailed investigation 
of environmental effects at the scale of galaxy groups and pairs. 

\begin{table*}
\caption {The average scaling relations for the four samples discussed in Sec.~3.}
\[
\label{scaletab}
\begin{array}{cccccccccccccccc}
\hline\hline
\noalign{\smallskip}
\multicolumn{4}{c}{\rm HI-normal~(Def_{HI}<0.5)} & & \multicolumn{3}{c}{\rm HI-deficient~(Def_{HI}\geq0.5)} & & \multicolumn{3}{c}{\rm Outside Virgo} & & \multicolumn{3}{c}{\rm Virgo} \\
\noalign{\smallskip}
x  &  <x>  &  <\log(M_{dust}/M_{*})> & N_{gal} & & <x>  &  <\log(M_{dust}/M_{*})> & N_{gal} & & <x>  &  <\log(M_{dust}/M_{*})> & N_{gal}& & <x>  &  <\log(M_{dust}/M_{*})> & N_{gal}\\
\noalign{\smallskip}
\hline
\log(M_{*}) &       8.63 &       -2.07 \pm       0.10  &          6 &  &             8.62 &	  -2.35 \pm	  0.11  &	   5 	&  &	   8.62 &	-2.19 \pm	0.11  & 	 6 &      &	 8.64 &       -2.21 \pm       0.13  &	       5 \\ 
 	    &	     9.08 &	  -2.17 \pm	  0.05  &	  47 &  &            9.03 &	  -2.68 \pm	  0.11  &	  16 	&  &	   9.06 &	-2.24 \pm	0.06  & 	41 &     &	 9.09 &       -2.42 \pm       0.10  &	      22 \\ 
 	     &       9.47 &	  -2.30 \pm	  0.04  &	  47 &  &            9.49 &	  -2.88 \pm	  0.10  &	  22 	&  &	   9.48 &	-2.39 \pm	0.07  & 	35 &     &	 9.46 &       -2.58 \pm       0.08  &	      34 \\ 
 	     &       9.99 &	  -2.77 \pm	  0.07  &	  36 &  &           10.03 &	  -3.43 \pm	  0.10  &	  35 	&  &	   9.99 &	-2.87 \pm	0.09  & 	36 &     &	10.04 &       -3.32 \pm       0.09  &	      35 \\ 
 	     &      10.51 &	  -3.01 \pm	  0.16  &	  19 &  &           10.48 &	  -3.91 \pm	  0.12  &	  33 	&  &	  10.46 &	-3.68 \pm	0.18  & 	21 &     &	10.51 &       -3.52 \pm       0.14  &	      31 \\ 
 	     &      11.00 &	  -3.11 \pm	  0.44  &	   3 &  &           10.94 &	  -4.11 \pm	  0.27  &	  11 	&  &	  10.91 &	-3.34 \pm	0.33  & 	 5 &     &	10.98 &       -4.21 \pm       0.31  &	       9 \\ 
\noalign{\smallskip}							     								 \noalign{\smallskip}					        
 NUV-r &       2.02 &       -2.07 \pm       0.04  &         28 &  &  	             2.14 &	  -2.29 \pm	  0.12  &	   3 	&  &	   2.03 &	-2.08 \pm	0.05  & 	23 &      &	  2.07 &       -2.12 \pm       0.08  &  	8 \\ 
  &	  2.46 &       -2.30 \pm       0.04  &         39 &  &  	             2.58 &	  -2.55 \pm	  0.10  &	   8 	&  &	   2.48 &	-2.37 \pm	0.04  & 	29 &      &	  2.47 &       -2.30 \pm       0.07  &         18 \\ 
  &	  2.98 &       -2.46 \pm       0.05  &         34 &  &  	             3.03 &	  -2.56 \pm	  0.08  &	  16 	&  &	   3.00 &	-2.44 \pm	0.06  & 	27 &      &	  2.99 &       -2.55 \pm       0.06  &         23 \\ 
  &	  3.50 &       -2.59 \pm       0.07  &         18 &  &  	             3.52 &	  -2.90 \pm	  0.07  &	  12 	&  &	   3.53 &	-2.68 \pm	0.07  & 	18 &      &	  3.47 &       -2.76 \pm       0.10  &         12 \\ 
  &	  3.98 &       -2.77 \pm       0.09  &         14 &  &  	             3.99 &	  -3.07 \pm	  0.16  &	  13 	&  &	   3.98 &	-2.87 \pm	0.20  & 	11 &      &	  3.99 &       -2.95 \pm       0.09  &         16 \\ 
  &	  4.45 &       -3.25 \pm       0.21  &  	7 &  &  	             4.48 &	  -3.26 \pm	  0.08  &	  16 	&  &	   4.52 &	-3.32 \pm	0.18  & 	 5 &      &	  4.46 &       -3.24 \pm       0.10  &         18 \\ 
  &	  5.01 &       -3.40 \pm       0.09  &  	4 &  &  	             5.04 &	  -3.51 \pm	  0.10  &	  18 	&  &	   5.01 &	-3.57 \pm	0.14  & 	10 &      &	  5.06 &       -3.43 \pm       0.10  &         12 \\ 
  &	  5.52 &       -4.21 \pm       0.30  &  	4 &  &  	             5.51 &	  -4.32 \pm	  0.10  &	  37 	&  &	   5.47 &	-4.14 \pm	0.16  & 	14 &      &	  5.53 &       -4.39 \pm       0.11  &         27 \\ 
\noalign{\smallskip}							     								 \noalign{\smallskip}					        
\log(\mu_{*}) &       7.12 &       -1.84 \pm       0.03  &         12 &  &           7.07 &	  -2.17 \pm	  0.11  &	   3 	&  &	   7.11 &	-1.86 \pm	0.05  & 	10 &      &	  7.11 &       -1.99 \pm       0.11  &  	5 \\ 
  &	  7.52 &       -2.16 \pm       0.03  &         58 &  &  	             7.50 &	  -2.53 \pm	  0.08  &	  18 	&  &	   7.54 &	-2.23 \pm	0.04  & 	44 &      &	  7.49 &       -2.28 \pm       0.05  &         32 \\ 
  &	  7.99 &       -2.46 \pm       0.03  &         45 &  &  	             7.98 &	  -3.00 \pm	  0.07  &	  36 	&  &	   7.97 &	-2.47 \pm	0.05  & 	38 &      &	  7.99 &       -2.90 \pm       0.06  &         43 \\ 
  &	  8.45 &       -2.95 \pm       0.06  &         36 &  &  	             8.52 &	  -3.70 \pm	  0.10  &	  46 	&  &	   8.50 &	-3.25 \pm	0.09  & 	42 &      &	  8.48 &       -3.50 \pm       0.11  &         40 \\ 
  &	  8.87 &       -3.80 \pm       0.32  &  	6 &  &  	             8.92 &	  -4.42 \pm	  0.14  &	  21 	&  &	   8.91 &	-4.07 \pm	0.25  & 	 9 &      &	  8.91 &       -4.39 \pm       0.16  &         18 \\ 
\hline
\hline
\end{array}
\]
\end{table*}

Contrary to the $M_{dust}/M_{*}$ vs. $M_{*}$ and $M_{dust}/M_{*}$ vs. $\mu_{*}$ relations, 
the relation between $M_{dust}/M_{*}$ vs. $NUV-r$ does not show any clear variation with environment 
or \hi\ content. Although gas-poor Virgo systems cover a larger range of colours and are offset towards redder 
colours than \hi-normal 'field' galaxies, all the four sub-samples here considered apparently lie on the same 
scaling relation at a $>$85\% significance level. This is different from what is observed in the case of the 
\hi\ scaling relations \citep{cortese11}, where also the HI-fraction vs. $NUV-r$ colour relation showed a 
variation with the environment. As we will see in the next sections, the independence of the $M_{dust}/M_{*}$ and $NUV-r$ on 
environment is consistent with a scenario in which dust is removed mainly when the environment is able to strip the ISM 
directly from the optical disk, thus reducing the star formation and moving a galaxy along the main relation. 
Being the $NUV-r$ colour a proxy for SSFR, our findings do not only confirm the recent results of \cite{dacunha10}, 
who pointed out the existence of a tight relation between star formation and dust mass in nearby galaxies. 
They also suggest that this relation might be a very useful tool for chemical evolution models, since it is apparently not affected by 
the environment. It is important to remember that our colours are not corrected for internal dust attenuation and therefore 
part of the scatter at red colours could easily be due to the fact that we are mixing truly passive objects with 
highly obscured ones. We plan to investigate in more detail the relation between total SFR and dust mass 
in a future paper, once accurate dust extinction corrections are developed. 

In summary, once combined our results indicate that, at fixed stellar mass and stellar mass surface 
density, the dust content of nearby galaxies varies as a function of environment and \hi\ content, although 
the effect of the environment on the dust seems to be weaker than in the case of the \hi.

Finally, since galaxy properties are also tightly linked to optical morphology (e.g., \citealp{roberts94}), 
it is interesting to investigate how the $M_{dust}/M_{*}$ ratio varies as a function of morphological type. 
As shown in Fig.~\ref{MDStypes}, the $M_{dust}/M_{*}$ ratio rapidly decreases when moving from 
late- to early-type galaxies (see also \citealp{skibba11}). A detailed analysis of the dust properties 
of early-type galaxies in the HRS is presented by \cite{smith12}. 
Not surprisingly, almost all the non-detections are early-type galaxies. 
Moreover, even at fixed morphological type, we find that \hi-deficient objects have a significantly smaller 
dust content than \hi-normal systems, reinforcing once more one of the main results emerging from this 
section: i.e., when the \hi\ is removed from a galaxy the dust content is affected as well.
This is even more evident when the mean $M_{dust}/M_{*}$ ratios per bin of morphological type and their standard deviations 
(i.e., the large rectangles in Fig.~\ref{MDStypes}) are considered. 

\begin{figure*}
  \centering
  \includegraphics[width=17cm]{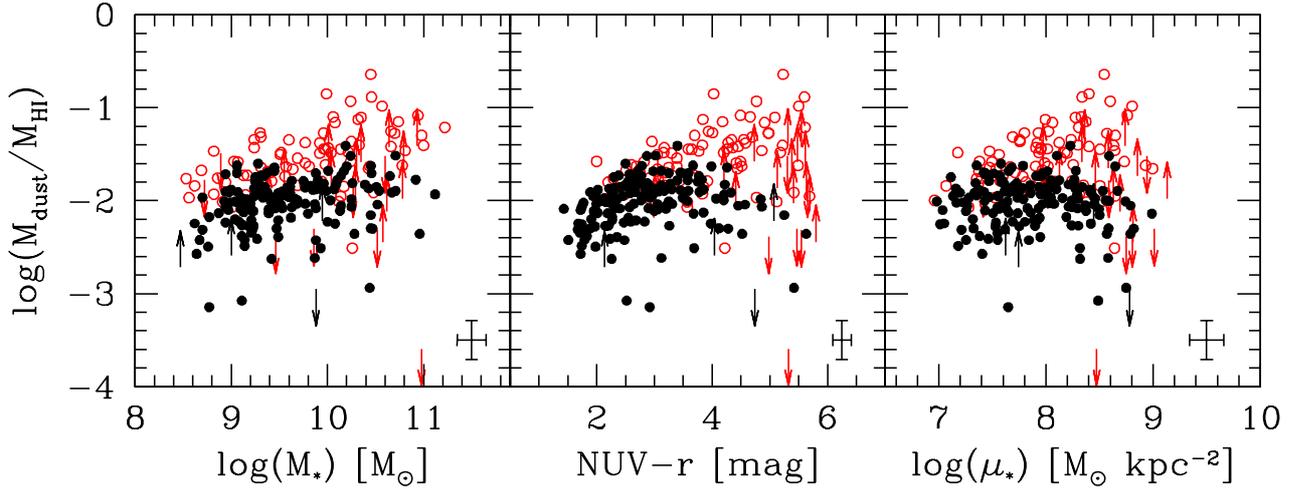}
     \caption{The dust-to-HI mass ratio as a function of stellar mass (right panel), $NUV-r$ colour (central panel) and stellar mass 
     surface density (left panel). Filled and empty circles indicate HI-normal and HI-deficient galaxies, respectively. Arrows show 
     upper limits in either dust or \hi\ mass. Typical errorbars are shown in the bottom-right corner of each panel.}
	 \label{MDgasall}
  \end{figure*}

\section{The dust-to-HI mass ratio}    
In the previous section we have shown that the $M_{dust}/M_{*}$ ratio follows scaling relations 
very similar to the ones observed for the HI-to-stellar mass ratio, suggesting that these two 
components of the ISM are tightly linked. In this section we investigate the variation 
of the $M_{dust}/M_{HI}$ ratio as a function of internal galaxy properties and environment 
following a similar approach as above. We focus on the separation between \hi-deficient 
and \hi-normal galaxies, since this turned out to be the ideal one to investigate environmental trends. 
 
In Fig.~\ref{MDgasall} we show the $M_{dust}/M_{HI}$ ratio as a function of stellar mass, $NUV-r$ 
colour and stellar mass surface density. Filled and empty symbols indicate \hi-normal and \hi-deficient 
galaxies, respectively. Arrows indicate both upper limits in \hi\ mass or in dust mass. Galaxies 
not detected at both radio and submm wavelengths are not shown.  
Contrary to what observed for the  $M_{dust}/M_{*}$ ratio, the dust-to-HI ratio varies less than 2 dex. 
Almost no correlation is found with the stellar mass surface density ($r=$0.1) while a weak correlation is 
found with stellar mass and $NUV-r$ colour ($r\sim$0.31 and dispersion along the y-axis $\sim$0.37 dex).
These relations become even weaker if we focus on \hi-normal galaxies. The $M_{dust}/M_{HI}$ vs. $M_{*}$ 
relation is still significant ($r\sim$0.28), showing that the $M_{dust}/M_{HI}$ ratio is an increasing 
function of stellar mass, as expected from the stellar mass vs. gas metallicity relation (e.g., \citealp{tremonti04}).
Conversely, the correlation coefficient for the $M_{dust}/M_{*}$ vs. $NUV-r$ relation drops to $r\sim$0.07.
This is mainly due to the presence of outliers at red colours ($NUV-r>$4), some of which 
are likely red just because they are highly attenuated and not because they are passive. 
If we limit our analysis to galaxies with $NUV-r<$4 the correlation coefficient 
increases to $\sim$0.4. 

\begin{figure}
  \centering
  \includegraphics[width=9cm]{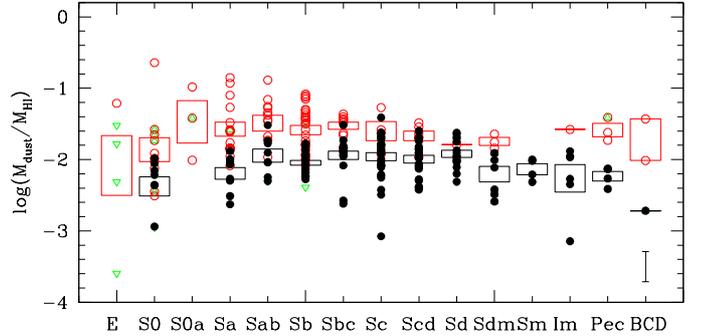}
     \caption{The $M_{dust}/M_{HI}$ ratios as a function of the morphological type. Symbols are as in Fig.~\ref{MDStypes}.}
	 \label{MDGtypes}
  \end{figure}

One of the most interesting results emerging from Fig.~\ref{MDgasall} is indeed the different behavior 
of \hi-deficient and \hi-normal galaxies.
The dynamic range covered by the $M_{dust}/M_{HI}$ ratio almost doubles when \hi-deficient 
galaxies are included into the picture. At fixed stellar mass, \hi-deficient objects have systematically 
higher $M_{dust}/M_{HI}$ ratios than \hi-normal galaxies. A similar offset is observed for the stellar mass 
surface density, although more pronounced at high $\mu_{*}$, whereas in the  $M_{dust}/M_{HI}$ vs. $NUV-r$ relation 
\hi-deficient galaxies appear mainly to extend the relation already seen for \hi-normal galaxies to redder colours, 
confirming that the gas removal and the reddening of the colours (i.e., quenching of the star formation) are related \citep{review,dEale,cortese09}.
This suggests that, when the gas is removed, the fraction of dust stripped from the galaxy is significantly lower 
than the \hi, thus automatically increasing the $M_{dust}/M_{HI}$ ratio in these objects. 
Similar results are obtained if we investigate the variation of the $M_{dust}/M_{HI}$ ratio as a function of morphological 
type (see Fig.~\ref{MDGtypes}). The $M_{dust}/M_{HI}$ ratio is almost independent of morphological type, showing 
(in the case of \hi-normal galaxies) a very small decrease only towards either early-type or irregulars/BCDs, consistent 
with what previously observed by \cite{draine07} for SINGS galaxies. At fixed morphological type, \hi-deficient galaxies 
have higher $M_{dust}/M_{HI}$ ratio, confirming that the dust is less affected by the environment than the gas. 
\begin{figure}
  \centering
  \includegraphics[width=9cm]{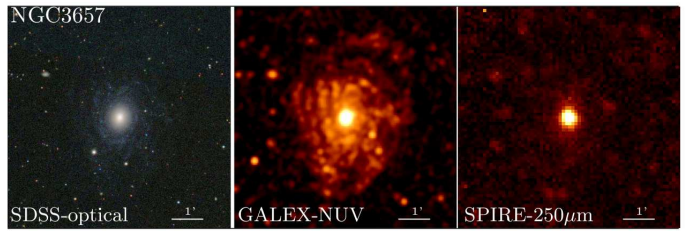}
     \caption{The extended UV disk of NGC3657. SDSS-optical (left), GALEX-NUV (center) and {\it Herschel}/SPIRE 250$\mu$m (right) images 
     are compared on the same spatial scale.}
	 \label{hrs52}
  \end{figure}

In addition to \hi-deficient galaxies showing a higher $M_{dust}/M_{HI}$ ratio than normal, it is interesting to note 
the presence of a few outliers having $M_{dust}/M_{HI}$ significantly lower ($\lesssim 10^{-3}$) than the average value 
for \hi-normal systems ($\sim10^{-2.1}$), namely: NGC3657, NGC4203, NGC4262, NGC4636 and UGC8045. 
In the case of UGC8045 and NGC4636 the low $M_{dust}/M_{HI}$ could just be due to a problem with the estimate of the \hi\ mass. 
UGC8045 could indeed be confused in \hi\ (i.e., more than one galaxy included in the Arecibo beam, \citealp{kent08}) and 
in the case of NGC4636 values published in the literature vary  
between $\sim10^{9}$ M$_{\odot}$ \citep{goldmine,knapp1978} and $<8\times10^{7}$ M$_{\odot}$ \citep{kumar83}.
Conversely, we can exclude a measurement problem for NGC3657, NGC4203 and NGC4262 and conclude that these systems 
have significantly higher amounts of atomic hydrogen compared to their dust content. 
This is confirmed when we look in detail at the properties of these systems. 
NGC4203 and NGC4262 have peculiar \hi\ morphologies \citep{krumm85,vandriel88} and several independent analyses have 
proposed that these galaxies might have recently accreted cold gas (see \citealp{cortese09} and references therein).
Although no \hi\ interferometric data are available for NGC3657, the huge ultraviolet extended disk revealed 
by GALEX images (see Fig.~\ref{hrs52}) suggests that this galaxy likely contains an extended \hi\ reservoir.
Thus, we speculate that the unusually low $M_{dust}/M_{HI}$ ratio observed in these systems may be due to the presence 
of a considerable amount of pristine gas not yet enriched by star formation, either because it has been recently accreted or 
because it has too low column density to form stars at high rate.
Unfortunately, no gas metallicity estimates for the outer parts of these galaxies are available. 
If confirmed, this result would suggest the use of the $M_{dust}/M_{HI}$ ratio 
as an alternative method to identify objects that have recently accreted a significant amount of pristine gas.

Finally, there is a caveat to this discussion; as discussed in Appendix A the relation between $M_{dust}/M_{HI}$ 
and stellar mass could become gradually weaker for $\beta<$2 (e.g., for $\beta=$1.5 the correlation 
coefficient of the $M_{dust}/M_{HI}$ vs. $M_{*}$ relation decreases to $\sim$0.27) and it might almost disappear 
in the extreme case of $\beta\sim$1.

\section{Discussion}
\subsection{The origin of the dust scaling relations}
We have shown that the dust-to-stellar mass ratio strongly anti-correlates with stellar mass, stellar mass 
surface density and colour over the whole dynamic range covered by the HRS, reproducing remarkably well similar 
scaling relations observed for the \hi-to-stellar mass ratio \citep{catinella10,cortese11}.
Similarly, the dust-to-stellar mass ratio monotonically decreases when moving from late- to early-type galaxies.
Although some of these scaling relations are not new [e.g., \cite{dacunha10} has recently shown that 
the $M_{dust}/M_{*}$ ratio strongly anti-correlates with SSFR], this is the first time that we can accurately quantify 
them for a volume-limited sample of galaxies spanning all morphologies and for which \hi\ information is available.

Once the prediction of dust formation and evolution models are taken into account 
(e.g., \citealp{dwek98,edmunds01,calura98}), it is quite easy to understand where these scaling relations are coming from. 
As nicely described by \cite{dacunha10}, the trends shown in Fig.~\ref{MDSall} can be seen as the result of the variation of 
SSFR with stellar mass (e.g., \citealp{boselli01,schiminovich07}).
Low mass, late-type systems are characterized by high SSFR and gas fraction. The gas can sustain the star formation activity and, 
consequently, a large fraction of dust is formed, likely exceeding the amount of dust grains destroyed in the ISM. 
When we move to higher stellar mass and earlier morphological types, the SSFR and gas fraction start decreasing and, as a result, 
the amount of dust produced will no longer be able to overcome those destroyed in the meantime. 
This would naturally explain why the dust-to-stellar mass ratio decreases with stellar mass, specific star formation rate, 
stellar mass surface density and when moving from irregulars to elliptical galaxies. In other words, what we are looking 
at is the change of the amount of baryonic mass stored in the ISM as a function of internal galaxy properties, 
explaining why \hi\ and dust follow the same scaling relations. 

\begin{figure}
  \centering
  \includegraphics[width=8cm]{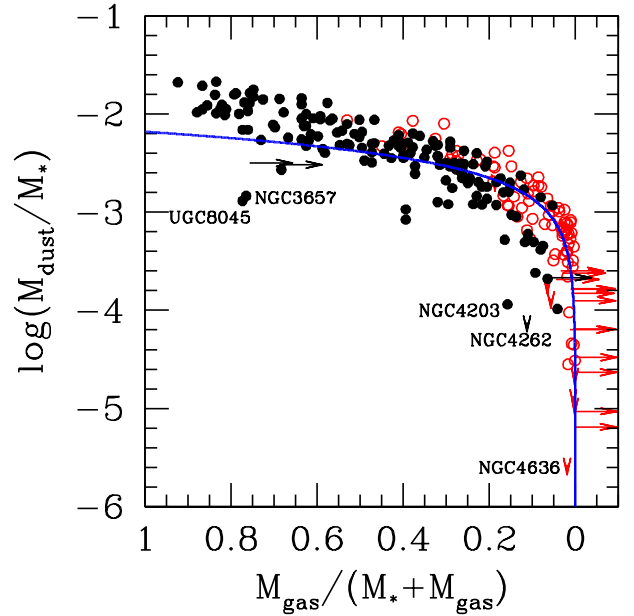}
     \caption{The dust-to-stellar mass ratio as a function of the total gas fractions for HRS galaxies. The solid blue line shows the predictions 
     of a simple closed-box models \citep{edmunds01}. See the text for details. The five galaxies
     with low dust-to-\hi\ ratio mass discussed 
     in Sec. 4 are highlighted. Symbols are as in Fig.~\ref{MDgasall}.}
	 \label{models}
  \end{figure}

To test this scenario qualitatively, we use the simple analytical approach developed by \cite{edmunds01} in the case of a closed-box. 
We use his Eq. 5 and 11 to determine how the dust-to-stellar mass ratio evolves as a function of the gas fraction. 
The model has six free parameters: namely, the efficiency of dust condensation from heavy elements made in 
stellar winds from massive stars and supernovae ($\chi_{1}$) and from Asymptotic Giant Branch stars ($\chi_{2}$), the effective 
yield ($p$), the fraction of mass of the ISM locked into stars ($\alpha$), the fraction of the ISM where mantles 
can grow ($\epsilon$) and the fraction of dust destroyed by star formation ($\delta$). Here we assume 
$\chi_{1}=$0.2 \citep{morgan03}, $\chi_{2}=$0.5 \citep{Zhukovska08}, $p=$ 0.012 and $\alpha=$0.80 (i.e., consistent with 
a \citealp{chabrier} IMF),  $\delta=$0.3 \citep{edmunds01} and $\epsilon=$0.5.
In order to estimate the total amount of gas of our galaxies from the \hi\ masses, we assume a 
molecular-to-atomic hydrogen gas ratio of 0.38 \citep{saintonge11} and include 
the contribution of helium and heavy elements (i.e., a factor 1.36).
The overall trend is not significantly affected if we account only for the contribution from 
atomic gas in the estimate of the gas fraction.  
We note that the goal of this exercise is just to establish whether the predictions of a simple 
closed-box model are roughly consistent with the shape of the trends we find. There are many free parameters 
and assumptions in the modeling, as well as a large uncertainty in the absolute value of our dust masses, 
thus looking for an exact fit does not seem meaningful at this stage. 
In addition, this is just a simple close-box model, whereas the evolutionary 
history of some galaxies in our sample (in particular early-type systems) might have been significantly different. 
The prediction of this simple model are shown in Fig.~\ref{models}. As expected, the $M_{dust}/M_{*}$ ratio decreases with the 
gas fraction confirming that the scaling relations we found are just a consequence of the tight link between 
dust and the cold gas in the ISM. Interestingly, the five galaxies with unusually low 
dust-to-\hi\ ratio, discussed in \S~4, are also outliers in Fig.~\ref{models}. Thus, for these objects, 
a closed-box approximation is likely inadequate. Our model is also 
unable to reproduce the very dust-rich systems in our sample. 
A similar tension between observations and models was recently found by \cite{dunne11} 
studying the properties of galaxies in the H-ATLAS survey. By comparing more detailed models predictions with observations 
they show how theoretical models are not able to reproduce galaxies with $\log(M_{dust}/M_{*})>-2.5$. The origin of this disagreement 
is still unclear, but we note that it may partially be solved if $\beta$ is lower than 2. 
We plan to investigate further this issue in the future when more accurate dust estimates will become possible.

\begin{figure*}
  \centering
  \includegraphics[width=15cm]{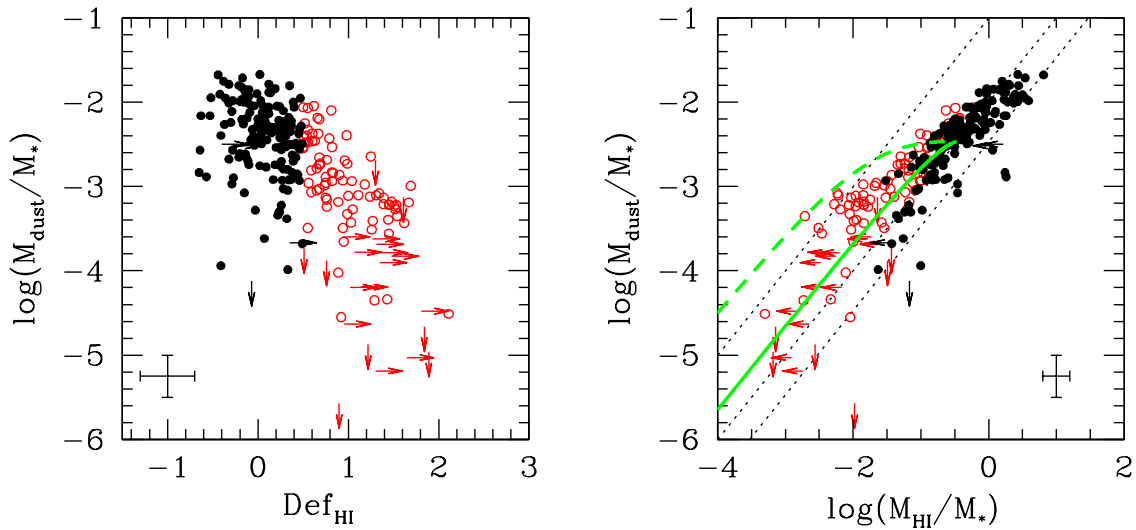}
     \caption{Left: The relation between dust-to-stellar mass and HI deficiency. 
     Right: The relation between dust-to-stellar mass and HI-to-stellar mass ratio.  
     The dotted lines indicate constant HI-to-dust ratios of (from left to right): 30-100-300. The green lines indicate 
     two different models of dust stripping assuming a constant (dashed) or exponential (solid) \hi\ surface density 
     profile. See text for details. Symbols are as in Fig.~\ref{MDgasall}. }
	 \label{dustdef}
  \end{figure*}


\subsection{Dust stripping}
The other important result emerging from our analysis is that at fixed stellar mass, stellar mass surface density 
and morphology, \hi-deficient/Virgo cluster galaxies have systematically lower $M_{dust}/M_{*}$ and 
higher $M_{dust}/M_{HI}$ ratios than \hi-normal/field systems. 
Once combined, these two independent results provide a consistent picture of the effects of the environment 
on the dust content of cluster galaxies.
On one side, the reduction in the  $M_{dust}/M_{*}$ ratio is a clear evidence for dust removal in cluster 
galaxies. This is reinforced by Fig.~\ref{dustdef} (left), where we show that the 
$M_{dust}/M_{*}$ ratio strongly decreases as a function of \hi\ deficiency (see also \citealp{corbelli11}).
On the other side, the higher $M_{dust}/M_{HI}$ ratio indicates that the effect of the environment 
on the dust content is significantly less dramatic compared to the amount of \hi\ stripped from the disk and 
dispersed into the intra-cluster medium. 
As already suggested by \cite{cortese10c}, this could be easily explained if the dust disk is significantly 
less extended than the gas disk \citep{thomas04} or, in other words, if the dust-to-gas ratio decreases 
monotonically with increasing distance from the center of galaxies \citep{bendo09,mateos09,pohlen10,magrini11}.

We test this interpretation in the right panel of Fig.~\ref{dustdef}, where we compare 
the strong correlation between $M_{dust}/M_{*}$ and $M_{HI}/M_{*}$ found in our data, with the 
predictions of a very simple toy model.
We assume that the stripping of the ISM is outside-in, with the outer parts being affected before the galaxy center, 
as expected in a ram-pressure stripping scenario \citep{koop06,review,n4569,cortese11}. 
Once the stripping starts, all the components of the ISM (i.e., both gas and dust) 
are completely removed from the disk up to the truncation radius. 
The dust mass surface density profile is exponential while 
we consider two different surface density distributions for the \hi . 
In the first one (solid line in Fig.~\ref{dustdef}), the \hi\ profile is exponential and the scalelength of the 
\hi\ disk is just 1.5 times the one of the dust disk \citep{thomas04}. In the second one (dashed line in Fig.~\ref{dustdef}), 
the \hi\ profile is constant \citep{bigiel10} and extends up to 8 times the scalelength of the dust disk. This corresponds to 
a \hi\ radius 1.5 times the optical/dust radius assuming that the dust follows the stellar distribution and that the 
ratio between optical scale-length and optical radius is $\sim$5.3 \citep{gav00}.
Both models are normalized so that the unperturbed galaxy has the typical $\log(M_{dust}/M_{*})$ of \hi-normal 
galaxies ($\sim-$2.46 dex) and $M_{dust}/M_{HI}=$100.  
We note that these two cases represent two extremes in the difference between the dust and \hi\ distribution in galaxies.
In Fig.~\ref{dustdef} (right) we show the evolution of the $M_{dust}/M_{*}$ and $M_{HI}/M_{*}$ predicted by the two models 
for different values of stripping radius. As expected, the range covered by these models reproduces the difference 
in $M_{dust}/M_{*}$ and $M_{HI}/M_{*}$ between \hi-normal and \hi-deficient galaxies. 
This supports our simple interpretation that the dust is affected by the environment in a less dramatic 
way than the \hi\ just because, like the molecular gas, it is more centrally concentrated than the atomic gas component. 
These results confirm once more dust stripping by environmental effects as an important mechanism for injecting dust 
grains into the intra-cluster medium, thus contributing to its metal enrichment \citep{popescu00}. 

The clear evidence of dust stripping emerging from our analysis leaves open the possibility that also the molecular 
hydrogen component in galaxies could be significantly perturbed in the center of clusters of galaxies. 
However, this issue is still highly debated \citep{boselligdust,fumagalli09} and, as clearly shown by our analysis, 
only a detailed comparison of molecular hydrogen properties for galaxies in different environment but with the same 
morphological type and stellar mass, will eventually allow us to quantify if and how molecules are directly removed 
from the star forming disks of infalling spirals.

\section{Summary}
In this paper, we have combined {\it Herschel}/SPIRE observations with optical, \hi\ and UV data to investigate 
the relation between dust content, galaxy properties and environment for the Herschel Reference Survey, 
a volume- magnitude- limited sample of $\sim$300 galaxies. 
Our main results are as follows:
\begin{itemize}

\item We find that the dust-to-stellar mass ratio strongly anti-correlates with stellar mass, stellar mass 
surface density and $NUV-r$ color across the whole range of parameters covered by our sample.
These relations are followed by all galaxies, regardless of environment or gas content, but Virgo/\hi-deficient 
galaxies show a systematically lower  dust-to-stellar mass at fixed stellar mass and morphological type.\\ 

\item Significantly weaker correlations are found between the dust-to-\hi\ mass ratio and internal galaxy properties, 
with the $M_{dust}/M_{HI}$ ratio mildly increasing with stellar mass, as expected from the mass-metallicity relation.
Gas-poor cluster galaxies have systematically higher dust content per unit of \hi\ mass than \hi-normal 
systems.\\

\item We show that the differences observed between \hi-deficient and \hi-normal galaxies provide strong evidence 
for dust stripping in cluster galaxies. However, we demonstrate that the fraction of dust removed from the disk is significantly 
lower than the \hi, and that this is likely just a consequence of the fact that the \hi\ disk is much more 
extended than the dust disk or, in other words, that the dust-to-atomic hydrogen ratio quickly declines in the outer regions.\\
 
\item Finally, we compare our results with the prediction of simple models of dust formation and evolution showing 
that the trends here presented are consistent with a simple picture in which the amount of dust in galaxies is regulated 
by the star formation activity and cold gas content.  
\end{itemize}

Although this work represents just the first step in the understanding of dust properties in galaxies, 
our investigation highlights the power of the HRS as an ideal local sample for galaxy evolution studies. 
The natural extension of the present analysis will be the study of the relation between dust properties, 
molecular hydrogen content and gas metallicity. Only after molecules and heavy elements in the ISM 
are included into the picture and more accurate dust mass estimates are available, will it be really possible 
to unveil the role of dust on galaxy evolution not only providing strong constraints for theoretical models 
but also a reference for high-redshift studies, i.e., the primary goal of a survey such as the HRS.

\begin{acknowledgements}
We thank all the people involved in the construction and the launch of {\it Herschel}. 
SPIRE has been developed by a consortium of institutes led
by Cardiff University (UK) and including Univ. Lethbridge
(Canada); NAOC (China); CEA, LAM (France); IFSI, Univ.
Padua (Italy); IAC (Spain); Stockholm Observatory (Sweden);
Imperial College London, RAL, UCL-MSSL, UKATC,
Univ. Sussex (UK); and Caltech, JPL, NHSC, Univ. Colorado
(USA). This development has been supported by national
funding agencies: CSA (Canada); NAOC (China);
CEA, CNES, CNRS (France); ASI (Italy); MCINN (Spain);
SNSB (Sweden); STFC and UKSA (UK); and NASA (USA). HIPE is a
joint development (are joint developments) by the Herschel
Science Ground Segment Consortium, consisting of ESA,
the NASA Herschel Science Center, and the HIFI, PACS
and SPIRE consortia.

GALEX is a NASA Small Explorer, launched in 2003 April. 
We gratefully acknowledge NASA's support for construction, operation and science analysis 
for the GALEX mission, developed in cooperation with the Centre National d'Etudes Spatiales (CNES) 
of France and the Korean Ministry of Science and Technology.

This publication makes use of data products from Two Micron All Sky Survey, 
which is a joint project of the University of Massachusetts and the Infrared Processing and Analysis 
Center/California Institute of Technology, funded by the National Aeronautics and Space 
Administration and the National Science Foundation.

Funding for the SDSS and SDSS-II has been provided by the Alfred P. Sloan Foundation, the Participating Institutions, 
the National Science Foundation, the U.S. Department of Energy, the National Aeronautics and Space Administration, 
the Japanese Monbukagakusho, the Max Planck Society, and the Higher Education Funding Council for England. 
The SDSS Web Site is http://www.sdss.org/.
The SDSS is managed by the Astrophysical Research Consortium for the Participating Institutions. 
The Participating Institutions are the American Museum of Natural History, Astrophysical Institute Potsdam, 
University of Basel, University of Cambridge, Case Western Reserve University, University of Chicago, 
Drexel University, Fermilab, the Institute for Advanced Study, the Japan Participation Group, Johns Hopkins University, 
the Joint Institute for Nuclear Astrophysics, the Kavli Institute for Particle Astrophysics and Cosmology, 
the Korean Scientist Group, the Chinese Academy of Sciences (LAMOST), Los Alamos National Laboratory, 
the Max-Planck-Institute for Astronomy (MPIA), the Max-Planck-Institute for Astrophysics (MPA), 
New Mexico State University, Ohio State University, University of Pittsburgh, University of Portsmouth, 
Princeton University, the United States Naval Observatory, and the University of Washington.

This research has made use of the NASA/IPAC Extragalactic Database (NED) which is operated 
by the Jet Propulsion Laboratory, California Institute of Technology, under contract 
with the National Aeronautics and Space Administration; and of the GOLDMine database. 

The research leading to these results has received funding from the European Community's Seventh Framework Programme 
(/FP7/2007-2013/) under grant agreement No 229517.

S. B., E. C., L. K. H., S. D. A., L. M. and C. P. acknowledge financial support by ASI through the ASI-INAF 
grant `HeViCS: the Herschel Virgo Cluster Survey' I/009/10/0. 

C. V. received support from the ALMA-CONICYT Fund for the Development of
Chilean Astronomy (Project 31090013) and from the Center of Excellence in
Astrophysics and Associated Technologies (PBF06).
\end{acknowledgements}

\begin{appendix}
\section{Testing the dust mass estimate based on SPIRE flux densities only.}
\begin{figure*}[!ht]
  \centering
  \includegraphics[width=15cm]{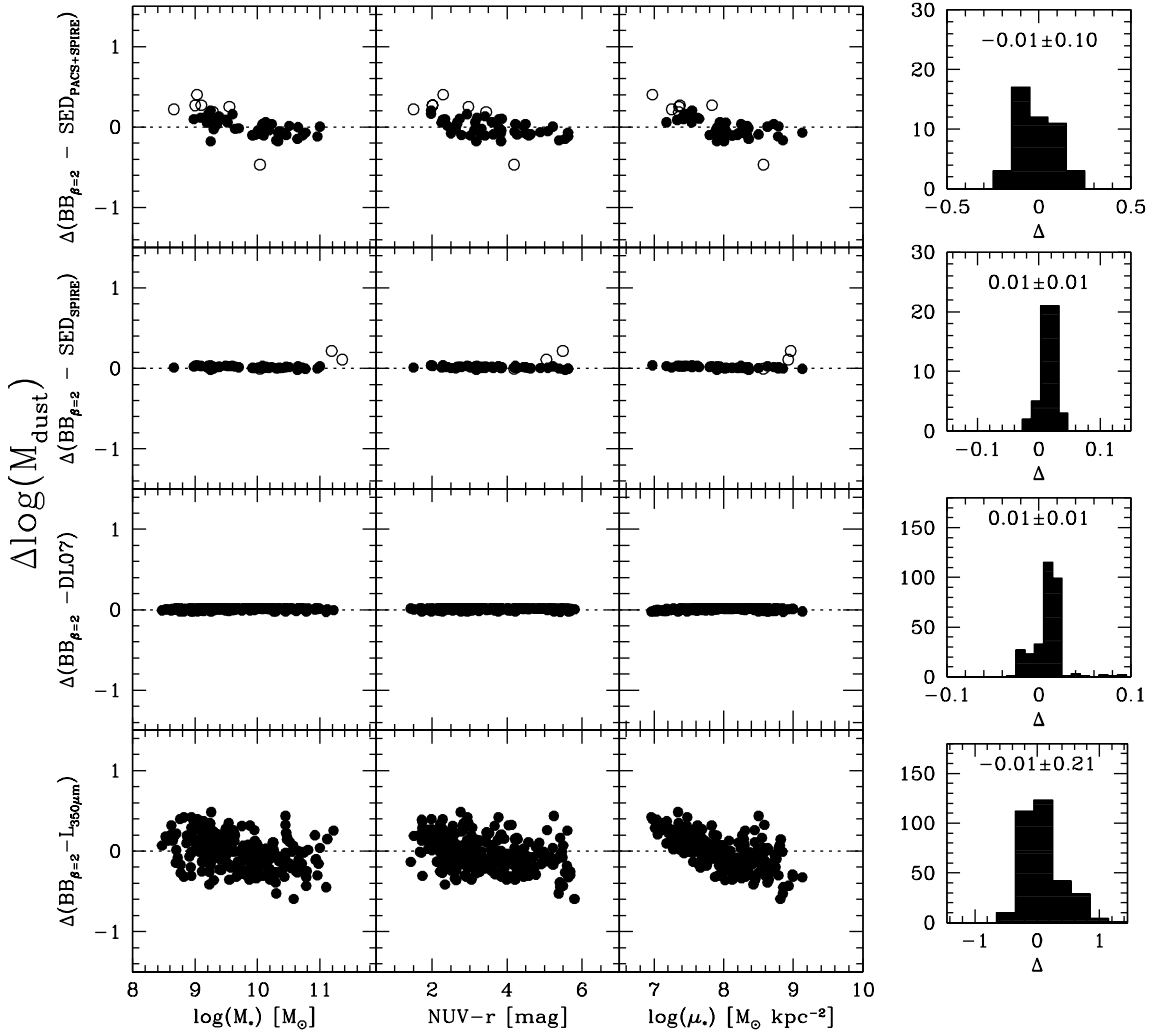}
     \caption{The logarithmic difference between the dust mass estimates presented in this work and those obtained from 
     modified black-body SED fitting of PACS+SPIRE (top row) and SPIRE-only (second row) data, using the SED library 
     of Draine \& Li (2007, third row) and using just the 350$\mu$m luminosity without any colour information (bottom row). 
     For each row, the right panel shows the histogram of the difference, its average value and standard deviation of the population.
     The empty circles in the top two rows indicate points for which the SED fitting routine provided a $\chi^{2}>$1.}
	 \label{test1}
  \end{figure*}
In this Appendix we investigate the reliability of the method described in \S~\ref{dustsec} to estimate 
dust masses. We proceed as follows. We compare our estimates with different 
methods all based on roughly the same assumptions on the properties of the underlying dust population 
(i.e., the value of $\beta$ and its variation with frequency). This is aimed at validating 
the methodology presented here. Then, we use our method for different values of $\beta$ to study 
how the assumption $\beta$=2 made in this paper could affect our results.
In both cases we are interested in the average difference and standard deviation between the various methods and 
in the presence of any possible systematic trend with stellar mass, stellar mass surface density and $NUV-r$ colour.

The results of the first step are shown in Fig.~\ref{test1}. In each row we show the difference between the logarithm 
of the dust mass obtained from two different methods as a function of stellar mass, $NUV-r$ colour and stellar mass surface density. 
In the right most panel we also present the histogram of the distribution of the difference, its average value and standard deviation. 

First of all, we compare our results with the ones obtained by \cite{davies11} for the 55 galaxies in common 
with our work (top row in Fig.~\ref{test1}). These are obtained using a single modified black-body SED fitting of the 
five PACS+SPIRE flux densities at 100, 160, 250, 350 and 500 $\mu$m and assuming the same dust emissivity 
as in this paper. In our analysis we only considered the points for which the SED fitting routine gives a $\chi^{2}<1$, 
but in Fig.~\ref{test1} we also show as empty circles the galaxies with $\chi^{2}>1$. We do not find any systematic difference in the average dust mass estimate and the 
scatter between the two methods is of the order of $\sim$0.1 dex, not significantly larger than the 
uncertainty of $\sim$0.08 dex in the dust mass estimate from the SED fitting technique \citep{davies11}.
Small systematic trends in the dust mass estimate are found as a function of stellar mass and stellar mass surface density. 
However, these are just marginal and within the error assumed throughout the paper.

Secondly, we consider the case of SED fitting of SPIRE-only points, following the same technique described above. 
Interestingly, we find a perfect agreement between the two methods and we could in theory just use the SED fitting 
of SPIRE fluxes only. The main reason why we decided not to do so is that the best fitting SED is significantly different 
from the one obtained by \cite{davies11} when combining PACS and SPIRE data since no constraint is provided below 250$\mu$m.
Therefore these are likely not a good representation of the far-infrared/submm emission for our galaxies and we preferred not 
to use them. Moreover, it is sometimes difficult to trust the uncertainty determined from the fitting technique since only 
based on the three SPIRE data.

Thirdly, we compare the estimate obtained from our method assuming a modified black-body SED and the one obtained from the 
same method but assuming the \cite{draine07b} SED (DL07) as described in \S~\ref{dustsec}. As shown in the third row of Fig.~\ref{test1} we find 
again a very good agreement between the two methods. This is due to the fact that, in the wavelength range covered by SPIRE, 
the SED predicted by the \cite{draine07b} models is very similar to a single temperature modified black-body 
with $\beta$=2 (see also \citealp{magrini11}).

\begin{figure}
  \centering
  \includegraphics[width=9cm]{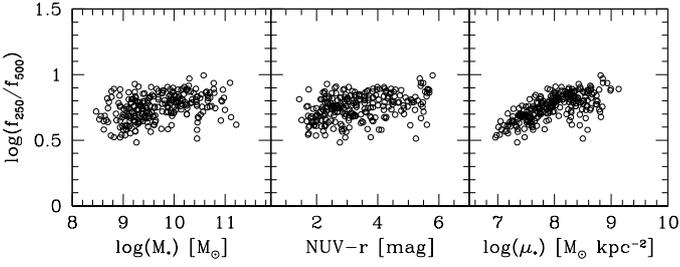}
     \caption{The 250-to-500 $\mu$m flux density ratio as a function of stellar mass (left), $NUV-r$ colour (center) and stellar mass 
     surface density (right).}
	 \label{colors}
  \end{figure}

\begin{figure}
  \centering
  \includegraphics[width=9.cm]{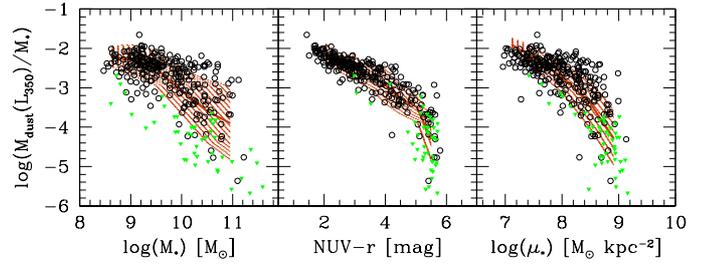}
     \caption{The dust-to-stellar mass ratio as a function of stellar mass (left), $NUV-r$ colour (center) and stellar mass 
     surface density (right), for dust masses estimated from the 350$\mu$m luminosities without taking into account any colour/temperature 
     variation in the sample. Symbols are as in Fig.~3. The orange lines and dashed regions show the average scaling 
     relations and $\pm$1$\sigma$ area presented in Fig.~3. }
	 \label{LDSall}
  \end{figure}

Finally, we consider the very simple case where only a submm flux density is available and no colour can be used 
(bottom row in Fig.~\ref{test1}). In details, we just assumed the same 250-to-500 $\mu$m ratio for all our galaxies 
(i.e., the mean value for the detections in our sample  $log(f_{250}/f_{500})=$0.75). 
This is done to investigate the role played by the colour term in our method and to warn  
how this very rough technique could affect scaling relations studies, in particular at high-redshift. 
As expected, this is by far the worse method to estimate dust masses showing a scatter of $\sim$0.2 dex. 
More importantly, we find systematic trends with stellar mass and stellar mass 
surface densities showing that, when compared with our method, this technique underestimates dust masses for low-mass galaxies and 
overestimate those of massive systems. The reason for these systematic trends is due to the fact that the 250-to-500 $\mu$m flux density 
ratio strongly correlates with stellar mass and stellar mass surface density (see Fig.~\ref{colors} and \citealp{boselli12}).
Thus not accounting for the different colours in our galaxies introduces systematic effects that can be seen in 
Fig.~\ref{LDSall}, where we compare some of the scaling relations discussed in this paper (orange dashed regions) with 
 those obtained using this method. While, as expected, the relation with 
 $NUV-r$ colour is not affected, the trends between dust-to-stellar mass ratio, stellar mass and stellar mass surface density change. 
 We thus suggest caution in using a technique based on only a single luminosity for a detailed quantification of 
 dust properties in galaxies. 
 
 In summary, these first tests confirm that our technique does not introduce any significant systematic effect, once the properties 
of the underlying dust component are fixed.

\begin{figure}
  \centering
  \includegraphics[width=8.cm]{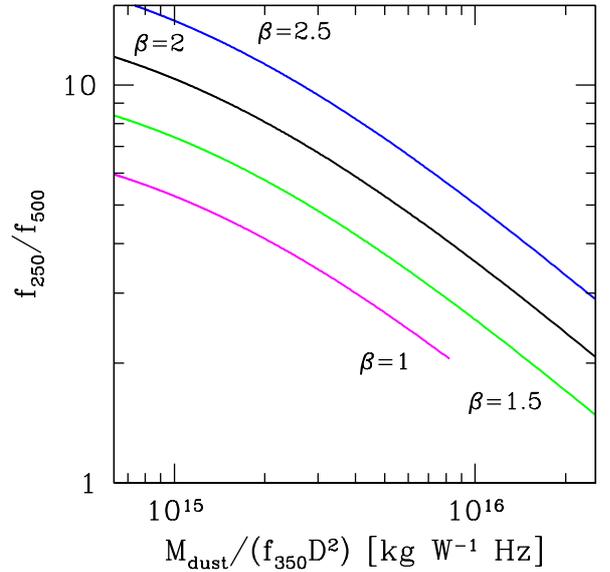}
     \caption{The relation between the ratio $M_{dust}/(f_{350} D^{2})$ and the 
250$\mu$m-to-500$\mu$m flux density ratio for a modified black-body SED with $\beta$=2.5 (blue), 2 (black), 1.5 (green) and 1 (magenta).}
	 \label{dustmass2}
  \end{figure}

\begin{figure*}
  \centering
  \includegraphics[width=15cm]{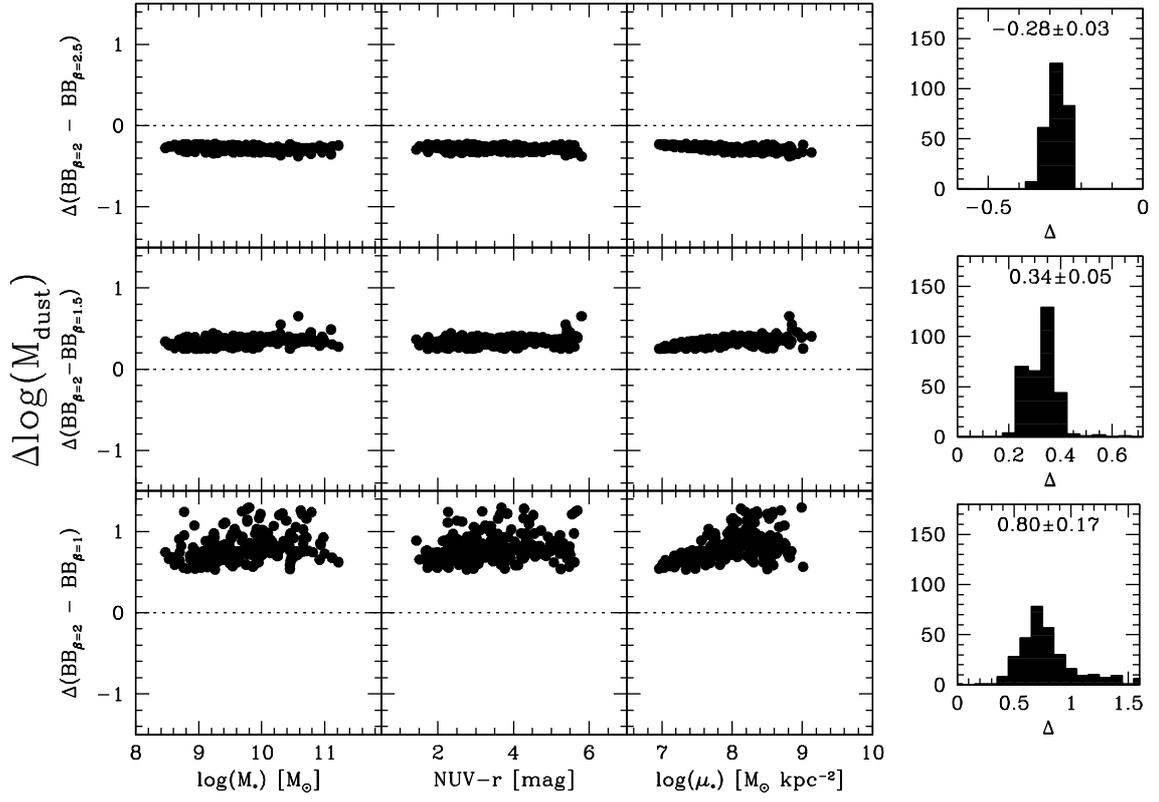}
     \caption{The logarithmic difference between the dust mass estimates presented in this work and the one obtained assuming 
     a modified black-body with $\beta$=2.5 (top row), 1.5 (middle row) and 1 (bottom row). For each row, the right panel shows the histogram 
     of the difference, its average value and standard deviation of the population.}
	 \label{test2}
  \end{figure*}

The next step is to see what happens if we change the value of $\beta$, keeping fixed the value of dust opacity 
at 350$\mu$m. In the following, we consider modified black-bodies 
with 1$\leq \beta \leq$2.5 which appear the most favoured range of values according to theoretical models and observations 
\citep{fink99,dupac03,gordon10,paradis10,plankbeta}.
It is important to note that, given the strong degeneracy between $\beta$ and dust temperature we had to modify the range 
of dust temperatures investigated in order to recover the range of submm colours observed in the HRS. 
In detail, while for $\beta$=2 and 2.5 we varied the temperature between 5 and 55 K, for $\beta$=1.5 and 1 we had 
to vary the temperature between 15 and 230 K. Moreover, the case $\beta$=1 is not able to reproduce the bluest 
colours observed in our sample (i.e., $f_{250}/f_{500}>$ 7.5, 37 galaxies in total) suggesting that this case is not 
a good representation of the dust SED for at least a fraction of our sample (see Fig.~\ref{dustmass2}).  
The results of our investigations are summarized in Fig.~\ref{test2}.
As expected, the absolute value of the dust mass varies with $\beta$ by almost a factor 10. 
Values of $\beta$ lower than 2 give dust masses smaller than the one used in this work, and viceversa. 
However, once this offset is taken into account, the intrinsic scatter in the dust mass estimate stays below $\sim$0.1 dex 
for 1.5$\leq \beta \leq$2.5 and goes up to just 0.17  dex for $\beta$=1 (for this case only those galaxies 
with $f_{250}/f_{500}\leq$ 7.5 are included), although with a significant tail of outliers. Moreover, very mild systematic trends 
are observed as a function of stellar mass, colour and stellar mass surface density but these could become important 
only for $\beta\sim 1$. As discussed in the text, this would reinforce the trend observed between the dust-to-stellar mass 
ratio but weaken the one between dust-to-\hi\ mass ratio and stellar mass. 
 

In conclusion, we can confidently state that the method presented here to estimate dust masses is a reliable solution  
when accurate SED fitting is not possible. The typical scatter with respect to our methods is of the order of $\sim$0.2 dex. 
Although different values of $\beta$ would imply different values of dust masses, the scaling relations 
discussed in this paper should remain valid at least for 1.5$\leq \beta \leq$2.5.
\end{appendix}

\begin{appendix}
\section{Recipes to estimate dust masses with SPIRE data only}
In this Appendix we provide the polynomial fit to the relations between the ratio $M_{dust}/(f_{350} D^{2})$ 
in units of kg W$^{-1}$ Hz and the 250$\mu$m-to-500$\mu$m flux density ratio which we used 
to estimate dust masses. We provide two different sets of coefficients obtained by convolving modified black-bodies characterized by 
different values of $\beta$ (i.e., 1, 1.5, 2 and 2.5) with the SPIRE RSRF for extended and point-like sources (Table~\ref{dusteq}).
Once the value $M_{dust}/(f_{350} D^{2})$ has been computed, the dust mass can be obtained from the following equation
\begin{equation}
\small
\log\big(\frac{M_{dust}}{M_{\odot}}\big) = 
\log \big( \frac{M_{dust}}{f_{350} D^{2}} \big) + 2\log\big(\frac{D}{Mpc}\big) + \log\big(\frac{f_{350}}{Jy}\big) -11.32
\end{equation}
We note that the difference degrees used to fit $\beta\geq$2 and $\beta<$2 are due to the different range of temperatures 
adopted for each case. For $\beta<$2 we had to use larger temperature range (15 to 230 K instead of 5 to 55 K) and, as a consequence, 
an higher order polynomial to fit the relation between the 250$\mu$m-to-500$\mu$m and $M_{dust}/(f_{350} D^{2})$ ratios.
The fits are optimized for the range of colours covered by our sample, i.e., 3$<f_{250}/f_{500}<$11.

\begin{table*}[!hb]
\caption {Relations to determine the $M_{dust}/(f_{350} D^{2})$ from the 250$\mu$m-to-500$\mu$m flux density ratio for extended sources 
and point sources.}
\[
\label{dusteq}
\begin{tabular}{ccccccccc}
\hline
\noalign{\smallskip}
\multicolumn{9}{c}{$\rm \log[M_{dust}/(f_{350} D^{2})]=a0+a1\times x+a2\times x^{2}+a3 \times x^{3}+ a4\times x^{4} + a5\times x^{5} + a6\times x^{6}$}\\
\noalign{\smallskip}
\hline
$\beta$ & x   &     a0   &  a1   &    a2  &    a3  &  a4 & a5 & a6  \\  
\noalign{\smallskip}									       
\hline											       
\noalign{\smallskip}	
\multicolumn{7}{c}{Point Sources}\\								     	  
1   &$\log(f_{250}/f_{500})$  &28.548  & -122.339  & 464.966  & -871.248  & 795.415  & -284.971  & - \\	  
1.5 &$\log(f_{250}/f_{500})$  &16.171  &  8.505  & -74.771   & 250.796   & -421.542  & 345.496  & -110.479 \\ 
2   &$\log(f_{250}/f_{500})$   &16.829  & -1.564  & 0.155   & -0.116   & -0.351 & -  & - \\
2.5 &$\log(f_{250}/f_{500})$   &17.053  & -1.615  & 0.163   &  0.085  & -0.347	& -  & - \\ 
\multicolumn{7}{c}{Extended Sources}\\								     	  
1   &$\log(f_{250}/f_{500})$  &31.928  & -149.111  & 546.406  & -987.384   & 871.581  & -302.422  & - \\	  
1.5 &$\log(f_{250}/f_{500})$  &15.857  & 13.672  & -100.986   & +313.155  & -495.517 & +386.798  & -118.623 \\ 
2   &$\log(f_{250}/f_{500})$  &16.880  & -1.559  & 0.160   & -0.079   & -0.363 & -  & - \\
2.5 &$\log(f_{250}/f_{500})$  &17.110  & -1.610  & 0.149   &  0.132   & -0.361  & -  & - \\ 

\noalign{\smallskip}
\hline
\end{tabular}
\]
\end{table*}

\end{appendix}
\end{document}